\documentclass[a4paper,11pt]{article}
\pdfoutput=1 

\usepackage{jheppub} 
\usepackage{mytikz}

\usepackage{subfig}
\usepackage{enumerate}

\newcommand{\bbS}{\mathbb{S}}
\newcommand{\bbR}{\mathbb{R}}
\newcommand{\D}{d}

\newcommand{\calO}{\mathcal{O}}

\newcommand{\AdS}{\mathrm{AdS}}

\newcommand{\sech}{\text{sech}}

\usepackage{bm}

\newtheorem{theorem}{Theorem}

\newtheorem{lemma}[theorem]{Lemma}

\newenvironment{argument}[1][Argument]{\noindent\textbf{#1.} }{\ \rule{0.5em}{0.5em}}

 \newcommand{\ket}[1]{|#1\rangle}

\usepackage{verbatim}

\title{Quantum tasks require islands on the brane}

\author[a]{Alex May}
\author[a]{and David Wakeham}

\affiliation[a]{Department of Physics and Astronomy, University of British Columbia
6224 Agricultural Road, Vancouver, B.C., V6T 1W9, Canada}

\emailAdd{may@phas.ubc.ca}
\emailAdd{daw@phas.ubc.ca}

\abstract{
In recent work, it was argued that quantum computations with inputs and outputs distributed in spacetime, or quantum tasks, impose constraints on entanglement in holographic theories. The resulting constraint was named the connected wedge theorem and can verified by a direct bulk proof using focusing arguments in general relativity. In this article we extend this work to the context of AdS/BCFT, where an end-of-the-world brane is present in the bulk. By considering quantum tasks which exploit information localized to the brane, we find a new connected wedge theorem. We apply this theorem to brane models of black holes, where it relates the formation of islands in the Ryu-Takayanagi formula to causal features of the ambient spacetime. In particular, we find that if the black hole interior is causally connected to the radiation system through the ambient spacetime, then an island forms. For constant tension branes in pure AdS the converse also holds.
}

\begin{document} 
\maketitle
\flushbottom

\section{Introduction}

In this article, we prove a theorem relating minimal surfaces and causal features of asymptotically AdS spacetimes which are ended by branes. Such geometries are relevant to the emergence of spacetime \cite{VanRaamsdonk2018,simidzija2020holo,may2020interpolating}, holographic approaches to cosmology \cite{Cooper2019, Antonini2019, hartman2020islands}, and the black hole information problem \cite{Rozali:2019day, Almheiri2020, Almheiri2020b,almheiri2020entanglement,geng2020information,geng2020massive,chen2020quantumPart1,chen2020quantumPart2}, where they model the formation of islands. In the island context, our theorem establishes that a causal connection from the black hole interior to the radiation system implies the existence of an island. 

Our work is motivated by the operational perspective on AdS/CFT initiated in \cite{may2019quantum} and elaborated in \cite{may2019holographic,may2021holographic}. In particular, the authors of \cite{may2019quantum,may2019holographic, may2021holographic} considered a quantum computation with inputs given at two boundary spacetime locations and outputs at two other boundary locations. Considering this computation from a bulk and boundary perspective leads to the \emph{connected wedge theorem}, a relationship between causal features of the bulk geometry and boundary entanglement. It was then possible to prove this theorem using tools from general relativity. 

Here we prove a similar result that applies specifically to the context of AdS spacetimes ended by branes. Such spacetimes are described by a manifold with boundary, along with a Lorentzian metric. The metric satisfies Einstein's equations along with a boundary condition set at the brane. Holographically, such spacetimes are proposed to be dual to conformal field theories with a boundary \cite{takayanagi2011holographic, Fujita2011}. We adapt the theorem of \cite{may2021holographic} to this setting.\footnote{The ``region based'' connected wedge theorem appearing in \cite{may2021holographic} is stronger than, and contains as a special case, the earlier theorem appearing in \cite{may2019quantum,may2019holographic}.} Our theorem is motivated again by an operational perspective on AdS/CFT, but involves considering quantum computations with one input location and two output locations. Additional information involved in the computation is localized to the brane. To distinguish our result from the earlier one we refer to it as the $1\rightarrow 2$ connected wedge theorem, and the earlier result of \cite{may2019holographic,may2021holographic} as the $2\rightarrow 2$ connected wedge theorem. 

\begin{figure}
    \centering
    \subfloat[\label{fig:connected-scattering}]{
    \tdplotsetmaincoords{15}{0}
    \begin{tikzpicture}[scale=1.4,tdplot_main_coords]
    \tdplotsetrotatedcoords{0}{30}{0}
    \draw[gray] (2,0,0) -- (2,6,0);
    
    \begin{scope}[tdplot_rotated_coords]
    
    \draw[thick,fill=black,opacity=0.4] (0,0,2) -- (0,0,-2) -- (0,6,-2) -- (0,6,2);
    
    \begin{scope}[canvas is xz plane at y=0]
    \draw[domain=-90:90,thick,gray] plot ({2*cos(\x)}, {2*sin(\x)});
    \end{scope}

    \begin{scope}[canvas is xz plane at y=6]
    \draw[domain=-90:90,thick,gray] plot ({2*cos(\x)}, {2*sin(\x)});
    \end{scope}
    
    \node at (1.7,2,1.7) {$\hat{\mathcal{V}}_1$};
    
    \draw[blue,domain=90:70,fill=blue,opacity=0.3] plot ({2*cos(\x)},{3+2*\x/90},{2*sin(\x)}) plot ({2*cos(\x)},{6.1-2*\x/90},{2*sin(\x)}) -- ({0},{3+2},{2});
    \draw[blue,domain=90:70,thick] plot ({2*cos(\x)},{3+2*\x/90},{2*sin(\x)}) plot ({2*cos(\x)},{6.1-2*\x/90},{2*sin(\x)}) -- ({0},{3+2},{2});

    \draw[blue,domain=90:70,fill=blue,opacity=0.5] plot ({2*cos(\x)},{3+2*\x/90},{-2*sin(\x)}) plot ({2*cos(\x)},{6.1-2*\x/90},{-2*sin(\x)}) -- ({0},{5},{-2});
    
    \node at (0.7,4.9,2) {$\hat{\mathcal{R}}_1$};
    \node at (0,4.75,-2.5) {$\hat{\mathcal{R}}_2$};
    
    \draw[thick, blue, opacity=0.3,domain=100:180] plot  ({0.62*sin(\x)},{4.5},{2+0.62*cos(\x)});
    
    \foreach \x in {90,...,180}
    {
    \draw[blue, opacity=0.2] ({0},{5},2) -- ({0.62*sin(\x)},{4.5},{2+0.62*cos(\x)});
    }
    
    \foreach \x in {90,...,180}
    {
    \draw[blue, opacity=0.2] ({0},{4.1},2) -- ({0.62*sin(\x)},{4.5},{2+0.62*cos(\x)});
    }
    
    \draw[thick, red] (2,1,0) -- (1,2,0);
    
    \begin{scope}[canvas is xz plane at y=2]
    \draw[domain=-90:90,thick,gray] plot ({2*cos(\x)}, {2*sin(\x)});
    \draw[domain=45:135,fill=lightgray,opacity=0.8] plot ({2*cos(\x-90)}, {2*sin(\x-90)}) -- (1.41,1.41) to [out=-135,in=0] (0,1) --  (0,-1) to [out=0,in=135] (1.41,-1.41) ;
    \draw[blue,thick] (1.41,1.41) to [out=-135,in=0] (0,1);
    \draw[blue,thick] (1.41,-1.41) to [out=135,in=0] (0,-1);
    \end{scope}
    
    \draw[thick,red,-triangle 45] (1,2,0) -- (0,3,0);
    
    \draw[thick, red,-triangle 45] (0,3,0) -- (0,4.5,-1.4);
    \draw plot [mark=*, mark size=1.5] coordinates{(0,4.5,-1.4)};
    
    \draw[thick, red,-triangle 45] (0,3,0) -- (0,4.5,1.4);
    \draw plot [mark=*, mark size=1.5] coordinates{(0,4.5,1.4)};
    
    \draw[thick, blue, opacity=0.3,domain=100:180] plot  ({0.62*sin(\x)},{4.5},{-2-0.62*cos(\x)});
    
    \foreach \x in {100,...,180}
    {
    \draw[blue, opacity=0.2] ({0},{5},-2) -- ({0.62*sin(\x)},{4.5},{-2-0.62*cos(\x)});
    }
    
    \foreach \x in {100,...,180}
    {
    \draw[blue, opacity=0.2] ({0},{4.1},-2) -- ({0.62*sin(\x)},{4.5},{-2-0.62*cos(\x)});
    }
    
    \draw[blue,thick,domain=135:230] plot ({2.5+0.86*cos(\x)},{1.4},{0.86*sin(\x)});
    
    \foreach \x in {135,...,230}
    {
    \draw[blue, opacity=0.2] ({2},{1.8},0) -- ({2.5+0.86*cos(\x)},{1.4},{0.86*sin(\x)});
    }
    
    \foreach \x in {135,...,230}
    {
    \draw[blue, opacity=0.2] ({2},{1},0) -- ({2.5+0.86*cos(\x)},{1.4},{0.86*sin(\x)});
    }
    
    \draw[thick,fill=blue,opacity=0.3] ({2},{1},0) -- ({2.5+0.86*cos(230)},{1.4},{0.86*sin(230)}) -- ({2},{1.8},0) -- ({2.5+0.86*cos(133)},{1.4},{0.86*sin(133)}) -- ({2},{1},0);
    
    \draw plot [mark=*, mark size=1.5] coordinates{(0,3,0)};
    \node[left] at (0,3,0) {$J^{\mathcal{E}}_{1\rightarrow 12}$};
    
    \foreach \x in {0,...,-90}
    {
    \draw[fill=black!50!,opacity=0.5] ({2*cos(\x/2)},{3+2*\x/180},{2*sin(\x/2)}) -- ({2*cos(\x/2)},{1-2*\x/180},{2*sin(\x/2)});
    }
    
    \foreach \x in {1,...,90}
    {
    \draw[fill=black!50!,opacity=0.5] ({2*cos(\x/2)},{3-2*\x/180},{2*sin(\x/2)}) -- ({2*cos(\x/2)},{1+2*\x/180},{2*sin(\x/2)});
    }
    
    \draw[dashed,domain=0:90] plot ({2*cos(\x)},{3+2*\x/90},{2*sin(\x)});
    \draw[dashed,domain=0:-90] plot ({2*cos(\x)},{3-2*\x/90},{2*sin(\x)});
    
    
    \draw[thick,domain=-45:0] plot ({2*cos(\x)},{3+2*\x/90},{2*sin(\x)});
    \draw[thick,domain=45:0] plot ({2*cos(\x)},{3-2*\x/90},{2*sin(\x)});
    
    \draw[thick,domain=-45:0] plot ({2*cos(\x)},{1-2*\x/90},{2*sin(\x)});
    \draw[thick,domain=45:0] plot ({2*cos(\x)},{1+2*\x/90},{2*sin(\x)});
    
    \end{scope}
    \end{tikzpicture}
    }
    \hfill
    \subfloat[\label{fig:disconnected-scattering}]{
    \tdplotsetmaincoords{15}{0}
    \begin{tikzpicture}[scale=1.4,tdplot_main_coords]
    \tdplotsetrotatedcoords{0}{30}{0}
    \draw[gray] (2,0,0) -- (2,6,0);
    
    \begin{scope}[tdplot_rotated_coords]
    
    \draw[thick,fill=black,opacity=0.4] (0,0,2) -- (0,0,-2) -- (0,6,-2) -- (0,6,2);
    
    \begin{scope}[canvas is xz plane at y=0]
    \draw[domain=-90:90,thick,gray] plot ({2*cos(\x)}, {2*sin(\x)});
    \end{scope}
    
    \node at (1.7,2,1.7) {$\hat{\mathcal{V}}_1$};
    
    \draw[blue,domain=90:70,fill=blue,opacity=0.3] plot ({2*cos(\x)},{3+2*\x/90-0.25},{2*sin(\x)}) plot ({2*cos(\x)},{6.1-2*\x/90-0.25},{2*sin(\x)}) -- ({0},{3+2-0.25},{2});
    \draw[blue,domain=90:70,thick] plot ({2*cos(\x)},{3+2*\x/90-0.25},{2*sin(\x)}) plot ({2*cos(\x)},{6.1-2*\x/90-0.25},{2*sin(\x)}) -- ({0},{3+2-0.25},{2});

    \draw[blue,domain=90:70,fill=blue,opacity=0.5] plot ({2*cos(\x)},{3+2*\x/90-0.25},{-2*sin(\x)}) plot ({2*cos(\x)},{6.1-2*\x/90-0.25},{-2*sin(\x)}) -- ({0},{5-0.25},{-2});
    
    \node at (0.7,4.9-0.25,2) {$\hat{\mathcal{R}}_1$};
    \node at (0,4.75-0.25,-2.5) {$\hat{\mathcal{R}}_2$};
    
    \draw[thick, blue, opacity=0.3,domain=90:180] plot  ({0.62*sin(\x)},{4.5-0.25},{2+0.62*cos(\x)});
    
    \foreach \x in {90,...,180}
    {
    \draw[blue, opacity=0.2] ({0},{5-0.25},2) -- ({0.62*sin(\x)},{4.5-0.25},{2+0.62*cos(\x)});
    }
    
    \foreach \x in {90,...,180}
    {
    \draw[blue, opacity=0.2] ({0},{4.1-0.25},2) -- ({0.62*sin(\x)},{4.5-0.25},{2+0.62*cos(\x)});
    }
    
    \draw[thick, red,-triangle 45] (1.55,1.6,0) -- (0,3.21,0);
    
    \begin{scope}[canvas is xz plane at y=2]
    \draw[domain=-90:90,thick,gray] plot ({2*cos(\x)}, {2*sin(\x)});
    \draw[domain=55:125,fill=lightgray,opacity=0.8] plot ({2*cos(\x-90)}, {2*sin(\x-90)}) -- (1.634,1.14) to [out=-140,in=+140] (1.634,-1.14);
    \draw[blue,thick] (1.634,1.14) to [out=-140,in=+140] (1.634,-1.14);
    \end{scope}
    
    \draw[dashed,domain=0:90] plot ({2*cos(\x)},{2.78+2*\x/90},{2*sin(\x)});
    \draw[dashed,domain=0:-90] plot ({2*cos(\x)},{2.78-2*\x/90},{2*sin(\x)});
    
    \draw[thick, blue, opacity=0.3,domain=90:180] plot  ({0.62*sin(\x)},{4.25},{-2-0.62*cos(\x)});
    
    \foreach \x in {90,...,180}
    {
    \draw[blue, opacity=0.2] ({0},{5-0.25},-2) -- ({0.62*sin(\x)},{4.25},{-2-0.62*cos(\x)});
    }
    
    \foreach \x in {90,...,180}
    {
    \draw[blue, opacity=0.2] ({0},{4.1-0.25},-2) -- ({0.62*sin(\x)},{4.25},{-2-0.62*cos(\x)});
    }
    
    \draw[blue,thick,domain=135:230] plot ({2.5+0.86*cos(\x)},{1.65},{0.86*sin(\x)});
    
    \foreach \x in {135,...,230}
    {
    \draw[blue, opacity=0.2] ({2},{2.05},0) -- ({2.5+0.86*cos(\x)},{1.65},{0.86*sin(\x)});
    }
    
    \foreach \x in {135,...,230}
    {
    \draw[blue, opacity=0.2] ({2},{1.25},0) -- ({2.5+0.86*cos(\x)},{1.65},{0.86*sin(\x)});
    }
    
    \draw[thick,fill=blue,opacity=0.3] ({2},{1.25},0) -- ({2.5+0.86*cos(230)},{1.65},{0.86*sin(230)}) -- ({2},{2.05},0) -- ({2.5+0.86*cos(133)},{1.65},{0.86*sin(133)}) -- ({2},{1.25},0);

    \foreach \x in {0,...,-70}
    {
    \draw[fill=black!50!,opacity=0.5] ({2*cos(\x/2)},{2.78+2*\x/180},{2*sin(\x/2)}) -- ({2*cos(\x/2)},{1.22-2*\x/180},{2*sin(\x/2)});
    }
    
    \foreach \x in {1,...,70}
    {
    \draw[fill=black!50!,opacity=0.5] ({2*cos(\x/2)},{2.78-2*\x/180},{2*sin(\x/2)}) -- ({2*cos(\x/2)},{1.22+2*\x/180},{2*sin(\x/2)});
    }
    
    \draw[thick, red,-triangle 45] (0,3.21,0) -- (0,5.21,-2);
    \draw[thick, red,-triangle 45] (0,3.21,0) -- (0,5.21,2);

    \draw plot [mark=*, mark size=1.5] coordinates{(0,3.21,0)};
    
    \draw[thick,domain=-35:0] plot ({2*cos(\x)},{2.78+2*\x/90},{2*sin(\x)});
    \draw[thick,domain=35:0] plot ({2*cos(\x)},{2.78-2*\x/90},{2*sin(\x)});
    
    \draw[thick,domain=-35:0] plot ({2*cos(\x)},{1.22-2*\x/90},{2*sin(\x)});
    \draw[thick,domain=35:0] plot ({2*cos(\x)},{1.22+2*\x/90},{2*sin(\x)});
    
    \begin{scope}[canvas is xz plane at y=6]
    \draw[domain=-90:90,thick,gray] plot ({2*cos(\x)}, {2*sin(\x)});
    \end{scope}
    
    \end{scope}
    \end{tikzpicture}
    }
    \caption{Illustration of Theorem \ref{thm:main}, shown with a zero tension brane. The input region is taken to be a point $\mathcal{C}_1=c_1$, while the output regions are the light blue half diamonds attached to the edge. The decision region $\hat{\mathcal{V}}_1$ is shown in black. a) When a boundary point $c_1$ and two edge points $r_1$, $r_2$ have a bulk scattering region which intersects the brane, the entanglement wedge of an associated domain of dependence (black shaded region) attaches to the brane. b) When there is no such scattering region, the entanglement wedge need not be connected.}
    \label{fig:phasetransition3d}
\end{figure}
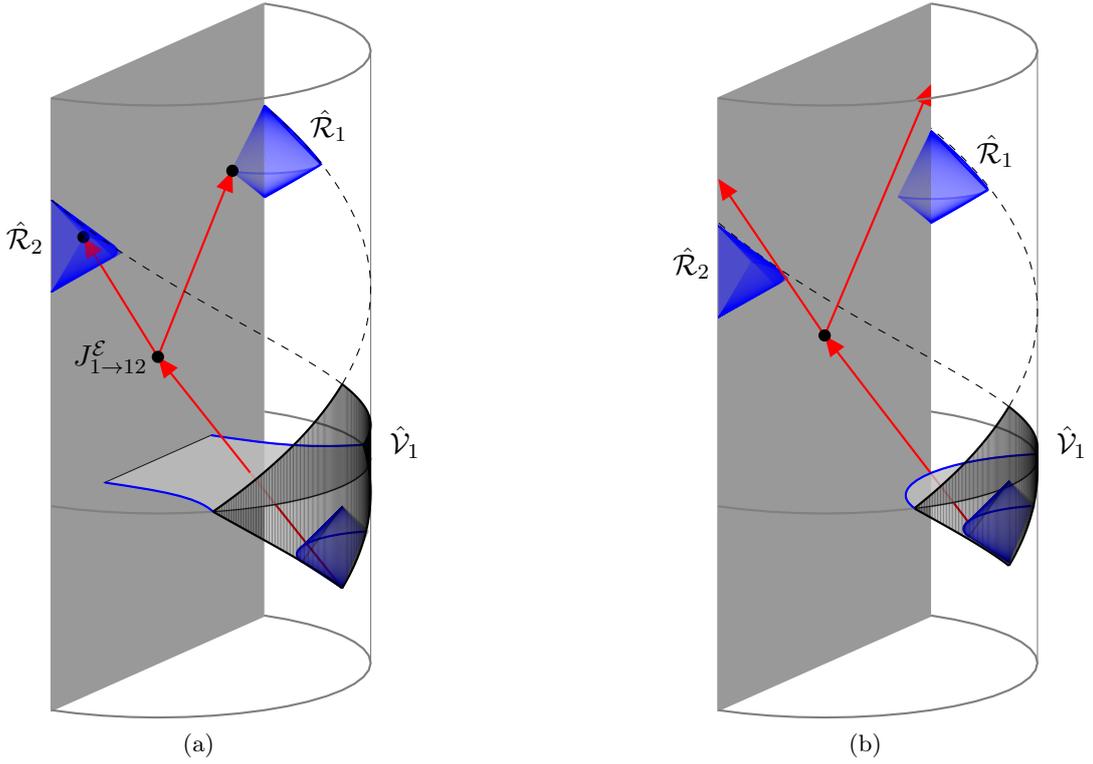

To state the $1\rightarrow 2$ connected wedge theorem more precisely, consider one `input' region $\hat{\mathcal{C}}_1$ and two `output' regions $\hat{\mathcal{R}}_1,\hat{\mathcal{R}}_2$. Note that we use hatted letters to denote regions in the boundary. We choose $\hat{\mathcal{R}}_1,\hat{\mathcal{R}}_2$ such that they touch the brane. The theorem is stated in terms of two additional spacetime regions constructed causally from $\hat{\mathcal{C}}_1,\hat{\mathcal{R}}_1,\hat{\mathcal{R}}_2$.

The first region is denoted $\hat{\mathcal{V}}_1$ and called the \emph{decision region}. It is defined by
\begin{align}\label{eq:definputregion}
    \hat{\mathcal{V}}_1 \equiv \hat{J}^+(\hat{\mathcal{C}}_1) \cap \hat{J}^-(\hat{\mathcal{R}}_1) \cap \hat{J}^-(\hat{\mathcal{R}}_2).
\end{align}
Here $\hat{J}^\pm(\hat{X})$ denotes the future and past of a region $\hat{X}$ taken in the boundary geometry. We will restrict our attention to choices of region $\hat{\mathcal{C}}_i,\hat{\mathcal{R}}_i$ such that $\hat{\mathcal{C}}_i\subseteq \hat{\mathcal{V}}_i$. 

The second region is called the \emph{entanglement scattering region} and is defined in the bulk spacetime. To define it, denote the entanglement wedge of a boundary region $\hat{X}$ by $X$, so that $X = \mathcal{E}_W(\hat{X})$. Further, denote the future and past of a region taken in the bulk geometry by $J^\pm(X)$. Then the entanglement scattering region is defined by
\begin{align}
    J_{1\rightarrow 12}^{\mathcal{E}} \equiv J^+(\mathcal{C}_1) \cap J^-(\mathcal{R}_1) \cap J^-(\mathcal{R}_2) \cap \mathcal{B},
\end{align}
where $\mathcal{B}$ denotes the end-of-the-world brane. This and definition \ref{eq:definputregion} are illustrated in figure \ref{fig:phasetransition3d}.

Our main result is as follows.
\begin{theorem}\label{thm:main}\textbf{($1\rightarrow 2$ connected wedge theorem)}
Consider three boundary regions $\hat{\mathcal{C}}_1,\hat{\mathcal{R}}_1,\hat{\mathcal{R}}_2$ in an asymptotically AdS$_{2+1}$ spacetime with an end-of-the-world brane. Require that $\hat{\mathcal{C}}_1\subseteq \hat{\mathcal{V}}_1$, and that $\hat{\mathcal{R}}_1,\hat{\mathcal{R}}_2$ touch the brane. Then if $J_{1\rightarrow 12}^{\mathcal{E}}$ is non-empty, the entanglement wedge of $\hat{\mathcal{V}}_1$ is attached to the brane. 
\end{theorem}
Note that in some cases $\hat{\mathcal{V}}_1$ may attach to the brane in the boundary, in which case the theorem is trivially true. The converse to this theorem does not hold, and we give an explicit example in the main text. 

To motivate our theorem, consider the following scenario. Suppose some classical information $q$ is encoded in the brane, either in the choice of boundary state or in brane-localized degrees of freedom. We leave unspecified at this stage in the argument where this corresponds to $q$ being localized in the boundary. Alice, an observer, receives a quantum state $H^q\ket{b}$ in region $\hat{\mathcal{C}}_1$. $H$ is the Hadamard operator, so if $q=0$ this is one of the states $\ket{0},\ket{1}$ and if $q=1$ this is one of the states $\ket{+},\ket{-}$. Without knowing $q$, Alice is not able to measure in the correct basis and learn $b$. However, Alice's goal is to bring $b$ to two regions $\hat{\mathcal{R}}_1$ and $\hat{\mathcal{R}}_2$, which will be attached to the CFT edge.


Causality requires that Alice can succeed in her task only when $q$ is stored in the patch of spacetime formed from the overlap of the past of $\hat{\mathcal{R}}_1, \hat{\mathcal{R}}_2$ (since she needs to send $b$ to both output regions) and the future of $\hat{\mathcal{C}}_1$ (since she needs the input $H^q\ket{b}$). We can consider this overlap in either the bulk or the boundary perspective. In the boundary we consider the future or past of the relevant boundary regions, $\hat{\mathcal{V}}_1=\hat{J}^+(\hat{\mathcal{C}}_1)\cap \hat{J}^-( \hat{\mathcal{R}}_1) \cap J^-(\hat{\mathcal{R}}_2)$. In the bulk perspective it is appropriate to consider the future or past of the corresponding entanglement wedges, $\mathcal{V}_1=J^+(\mathcal{C}_1)\cap J^-(\mathcal{R}_1) \cap J^-(\mathcal{R}_2)$. In either case if the overlap contains $q$, Alice can complete her task. 

When the bulk overlap $\mathcal{V}_1$ intersects the brane it contains $q$. This is the just the statement that $J^\mathcal{E}_{1\rightarrow 12}= \mathcal{V}_1\cap \mathcal{B}$ is non-empty. Then in the bulk picture Alice can complete her goal of bringing $b$ to $\mathcal{R}_1$ and $\mathcal{R}_2$. Meanwhile, for a holographic BCFT, entanglement wedge reconstruction tells us that the information stored in $\hat{\mathcal{V}}_1$ is geometrized as bulk degrees of freedom in its entanglement wedge. Thus the entanglement wedge of $\hat{\mathcal{V}}_1$ should include the brane whenever $J^\mathcal{E}_{1\rightarrow 12}$ is non-empty, which is the claim of the theorem.


In earlier work \cite{may2019quantum,may2019holographic}, the case where input and output regions consisted of single points was considered. For the applications of our theorem to islands discussed in section \ref{sec:islands} a point based version of Theorem \ref{thm:main} is appropriate. Partly this is because the statement in terms of regions is equivalent to the point based statement in the setting of pure AdS with an ETW brane. When not considering pure AdS spacetimes however the region based statement is stronger, so we have included the statement and proof of the more general theorem.

Notice that if we choose $\hat{\mathcal{C}}_i$ such that $\hat{\mathcal{C}}_i=\hat{\mathcal{V}}_i$ the theorem is not useful, in that finding the entanglement scattering region already involves determining the entanglement wedge of $\hat{\mathcal{V}}_i$. In this case however we can consider the minimal extremal surface which is not attached to the brane, call it $\gamma'_{{\mathcal{V}}_1}$, and define the region $W[\gamma'_{{\mathcal{V}}_1}]$ whose boundary is $\gamma'_{{\mathcal{V}}_1}\cup {\mathcal{V}}_1$. Note that $W[\gamma'_{{\mathcal{V}}_1}]$ is contained within the true entanglement wedge $W[\gamma_{{\mathcal{V}}_1}]$, which in general may be larger. In fact it will be larger whenever the minimal extremal surface $\gamma_{{\mathcal{V}}_1}$ is attached to the brane. Then Theorem \ref{thm:main} can be used to conclude that if 
\begin{align}
    (J')^{\mathcal{E}}_{1\rightarrow 12} \equiv J^+(W[\gamma'_{{\mathcal{V}}_1}]) \cap J^-(\mathcal{R}_1) \cap J^-(\mathcal{R}_2) \cap \mathcal{B} \neq \varnothing
\end{align}
then $W[\gamma'_{{\mathcal{V}}_1}]$ will not be the full entanglement wedge, and instead it is the brane attached extremal surface which is minimal.

\subsection*{Outline of paper}

We now give a summary of this paper. 

In section \ref{sec:background} we review the AdS/BCFT correspondence, which gives a holographic dual description of asymptotically AdS spacetimes with ETW branes. 

In section \ref{sec:QIargument}, we give the quantum information based argument for the $1\rightarrow 2$ theorem. The basic structure of the argument is to consider the regions $\{\hat{\mathcal{C}}_1,\hat{\mathcal{R}}_1,\hat{\mathcal{R}}_2\}$ as the spacetime locations for inputs and outputs to a quantum computation. We consider a particular such `distributed quantum computation', or \emph{quantum task}, in order to argue for the necessity of large correlation between the decision region $\hat{\mathcal{V}}_1$ and the brane. In brief, completing the task will be possible in the bulk perspective whenever $J_{1\rightarrow 12}^{\mathcal{E}}$ is non-empty, while completing it in the boundary will be possible whenever region $\hat{\mathcal{V}}_1$ knows information stored on the brane. We argue the task being completed in the bulk implies it is completed in the boundary. To ensure this is possible, a portion of the brane must be in the entanglement wedge of $\hat{\mathcal{V}}_1$.

In section \ref{sec:GRproof}, we prove the $1\rightarrow 2$ theorem from the bulk gravity perspective. The proof is by contradiction: we begin by assuming the scattering region is non-empty and the Ryu-Takayanagi surface $\tilde{\gamma}_{{\mathcal{V}}_1}$ enclosing the entanglement wedge of $\hat{\mathcal{V}}_1$ takes on the brane-detached configuration. Then, we show that this leads to the existence of a surface called the null membrane, which connects $\tilde{\gamma}_{{\mathcal{V}}_1}$ to a brane-attached surface of less area, so that the candidate surface $\tilde{\gamma}_{{\mathcal{V}}_1}$ cannot have been the correct one. 

In section \ref{sec:constantT}, we study bulk gravity solutions in $2+1$ dimensions that have a constant tension brane and are locally pure AdS. We verify the theorem explicitly for those solutions by comparing calculations of the entanglement entropy with features of bulk null geodesics. We also find that the converse to the theorem holds for these solutions.

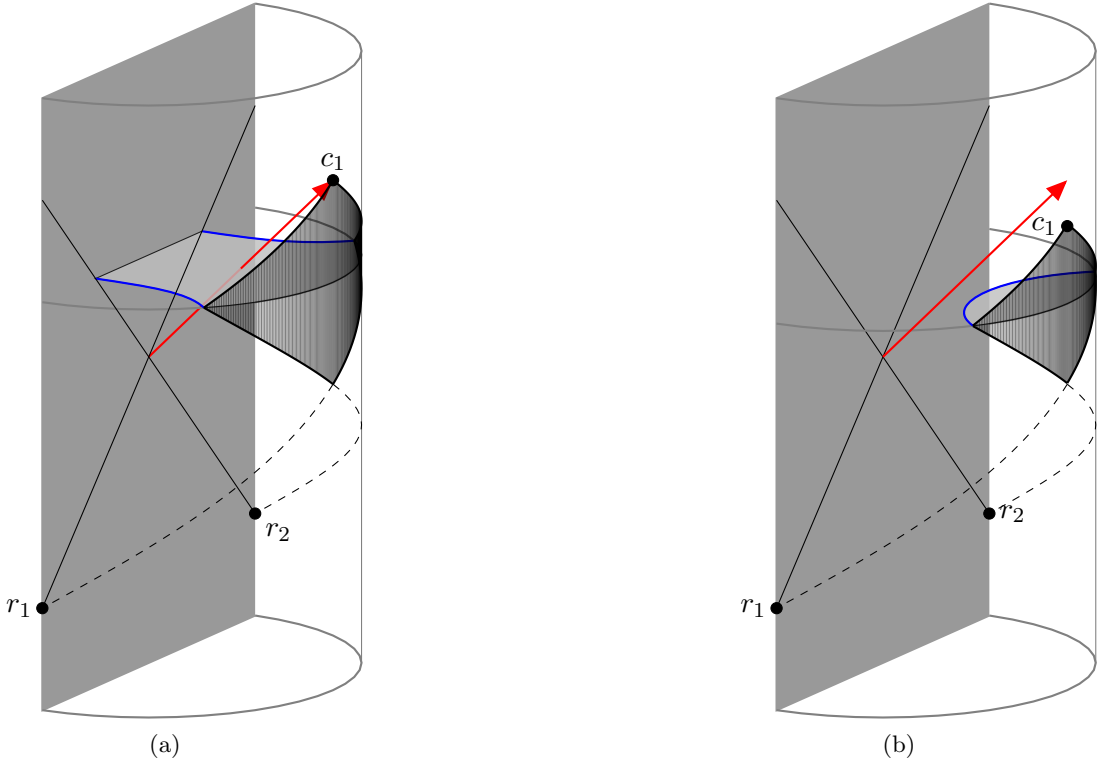
\begin{figure}
    \centering
    \subfloat[\label{fig:connected-scatteringflipped}]{
    \tdplotsetmaincoords{15}{0}
    \begin{tikzpicture}[scale=1.4,tdplot_main_coords]
    \tdplotsetrotatedcoords{0}{30}{0}
    \draw[gray] (2,0,0) -- (2,6,0);
    
    \begin{scope}[tdplot_rotated_coords]
    
    \draw[thick,fill=black,opacity=0.4] (0,0,2) -- (0,0,-2) -- (0,6,-2) -- (0,6,2);
    
    \begin{scope}[canvas is xz plane at y=0]
    \draw[domain=-90:90,thick,gray] plot ({2*cos(\x)}, {2*sin(\x)});
    \end{scope}

    \begin{scope}[canvas is xz plane at y=6]
    \draw[domain=-90:90,thick,gray] plot ({2*cos(\x)}, {2*sin(\x)});
    \end{scope}
    
    \draw[thick, red] (0,3,0) -- (1,4,0);
    
    \begin{scope}[canvas is xz plane at y=4]
    \draw[domain=-90:90,thick,gray] plot ({2*cos(\x)}, {2*sin(\x)});
    \draw[domain=45:135,fill=lightgray,opacity=0.8] plot ({2*cos(\x-90)}, {2*sin(\x-90)}) -- (1.41,1.41) to [out=-135,in=0] (0,1) --  (0,-1) to [out=0,in=135] (1.41,-1.41) ;
    \draw[blue,thick] (1.41,1.41) to [out=-135,in=0] (0,1);
    \draw[blue,thick] (1.41,-1.41) to [out=135,in=0] (0,-1);
    \end{scope}
    
    \draw (0,1,2) -- (0,5,-2);
    \draw (0,1,-2) -- (0,5,2);
    
    \draw[thick, red,-triangle 45] (1,4,0) -- (2,5,0);
    
    \draw plot [mark=*, mark size=1.5] coordinates{(2,5,0)};
    \node[above] at (2,5,0) {$c_1$};
    
    \draw plot [mark=*, mark size=1.5] coordinates{(0,1,-2)};
    \node[left] at (0,1,-2) {$r_1$};
    
    \draw plot [mark=*, mark size=1.5] coordinates{(0,1,2)};
    \node[below right] at (0,1,2) {$r_2$};
    
    \foreach \x in {0,...,-90}
    {
    \draw[fill=black!50!,opacity=0.5] ({2*cos(\x/2)},{5+2*\x/180},{2*sin(\x/2)}) -- ({2*cos(\x/2)},{3-2*\x/180},{2*sin(\x/2)});
    }
    
    \foreach \x in {1,...,90}
    {
    \draw[fill=black!50!,opacity=0.5] ({2*cos(\x/2)},{5-2*\x/180},{2*sin(\x/2)}) -- ({2*cos(\x/2)},{3+2*\x/180},{2*sin(\x/2)});
    }
    
    \draw[dashed,domain=0:90] plot ({2*cos(\x)},{3-2*\x/90},{2*sin(\x)});
    \draw[dashed,domain=0:-90] plot ({2*cos(\x)},{3+2*\x/90},{2*sin(\x)});
    
    
    \draw[thick,domain=-45:0] plot ({2*cos(\x)},{5+2*\x/90},{2*sin(\x)});
    \draw[thick,domain=45:0] plot ({2*cos(\x)},{5-2*\x/90},{2*sin(\x)});
    
    \draw[thick,domain=-45:0] plot ({2*cos(\x)},{3-2*\x/90},{2*sin(\x)});
    \draw[thick,domain=45:0] plot ({2*cos(\x)},{3+2*\x/90},{2*sin(\x)});
    
    \end{scope}
    \end{tikzpicture}
    }
    \hfill
    \subfloat[\label{fig:disconnected-scatteringflipped}]{
    \tdplotsetmaincoords{15}{0}
    \begin{tikzpicture}[scale=1.4,tdplot_main_coords]
    \tdplotsetrotatedcoords{0}{30}{0}
    \draw[gray] (2,0,0) -- (2,6,0);
    
    \begin{scope}[tdplot_rotated_coords]
    
    \draw[thick,fill=black,opacity=0.4] (0,0,2) -- (0,0,-2) -- (0,6,-2) -- (0,6,2);
    
    \begin{scope}[canvas is xz plane at y=0]
    \draw[domain=-90:90,thick,gray] plot ({2*cos(\x)}, {2*sin(\x)});
    \end{scope}

    \begin{scope}[canvas is xz plane at y=6]
    \draw[domain=-90:90,thick,gray] plot ({2*cos(\x)}, {2*sin(\x)});
    \end{scope}
    
    \draw (0,1,2) -- (0,5,-2);
    \draw (0,1,-2) -- (0,5,2);
    
    \begin{scope}[canvas is xz plane at y=3.79]
    \draw[domain=-90:90,thick,gray] plot ({2*cos(\x)}, {2*sin(\x)});
    \draw[domain=55:125,fill=lightgray,opacity=0.8] plot ({2*cos(\x-90)}, {2*sin(\x-90)}) -- (1.634,1.14) to [out=-140,in=+140] (1.634,-1.14);
    \draw[blue,thick] (1.634,1.14) to [out=-140,in=+140] (1.634,-1.14);
    \end{scope}
    
    \draw[dashed,domain=0:90] plot ({2*cos(\x)},{3-2*\x/90},{2*sin(\x)});
    \draw[dashed,domain=0:-90] plot ({2*cos(\x)},{3+2*\x/90},{2*sin(\x)});
    
    \foreach \x in {0,...,-70}
    {
    \draw[fill=black!50!,opacity=0.5] ({2*cos(\x/2)},{2.78+2*\x/180+1.79},{2*sin(\x/2)}) -- ({2*cos(\x/2)},{1.22-2*\x/180+1.79},{2*sin(\x/2)});
    }
    
    \foreach \x in {1,...,70}
    {
    \draw[fill=black!50!,opacity=0.5] ({2*cos(\x/2)},{2.78-2*\x/180+1.79},{2*sin(\x/2)}) -- ({2*cos(\x/2)},{1.22+2*\x/180+1.79},{2*sin(\x/2)});
    }
    
    
    \draw plot [mark=*, mark size=1.5] coordinates{(2,4.55,0)};
    \node[left] at (2,4.55,0) {$c_1$};
    
    \draw plot [mark=*, mark size=1.5] coordinates{(0,1,-2)};
    \node[left] at (0,1,-2) {$r_1$};
    
    \draw plot [mark=*, mark size=1.5] coordinates{(0,1,2)};
    \node[right] at (0,1,2) {$r_2$};
    
    \draw[thick,domain=-35:0] plot ({2*cos(\x)},{2.78+2*\x/90+1.79},{2*sin(\x)});
    \draw[thick,domain=35:0] plot ({2*cos(\x)},{2.78-2*\x/90+1.79},{2*sin(\x)});
    
    \draw[thick,domain=-35:0] plot ({2*cos(\x)},{1.22-2*\x/90+1.79},{2*sin(\x)});
    \draw[thick,domain=35:0] plot ({2*cos(\x)},{1.22+2*\x/90+1.79},{2*sin(\x)});
    
    \draw[thick, red,-triangle 45] (0,3,0) -- (2,5,0);
    
    \end{scope}
    \end{tikzpicture}
    }
    \caption{Theorem \ref{thm:main} along with time reversal implies a $2\rightarrow 1$ connected wedge theorem. We can view the light rays beginning at $r_1$ and $r_2$ as defining the horizons of a black hole. The region $\hat{\mathcal{V}}_1$ is then the radiation system. (a) When a light ray reaches $\hat{\mathcal{V}}_1$ from the black hole interior, the entanglement wedge of $\hat{\mathcal{V}}_1$ must connect to the brane, so that $\hat{\mathcal{V}}_1$ reconstructs a portion of the interior. (b) When the black hole is causally disconnected from the black hole interior, the entanglement wedge of $\hat{\mathcal{V}}_1$ may be disconnected from the brane.}
    \label{fig:Islandappearance}
\end{figure}

In section \ref{sec:islands}, we take up the discussion of islands. We study in particular island formation in BCFT models of black holes, following the set-up of \cite{Rozali:2019day} closely (see also \cite{Almheiri2020, Almheiri2020b,almheiri2020entanglement,geng2020information,geng2020massive,chen2020quantumPart1,chen2020quantumPart2}). In that context the brane is the black hole, the CFT boundary is the dual quantum mechanical description of the black hole, and the CFT is the bath system, into which information from the black hole may escape. We apply the time reversed statement of Theorem \ref{thm:main} to this setting, and take the input and out regions to be points $\hat{\mathcal{C}}_1=c_1$, $\hat{\mathcal{R}}_1=r_1$, $\hat{\mathcal{R}}_2=r_2$. In doing so we find that the point $c_1$, now at some late time, controls the time for which Hawking radiation has been collected from the black hole. Meanwhile the points $r_1$ and $r_2$ define the black hole event horizons. Moving $c_1$ to gradually later times, and so collecting more Hawking quanta, the scattering region $J_{12\rightarrow 1}^{\mathcal{E}}$ opens, which now corresponds to $c_1$ coming into causal contact with the black hole interior. Theorem \ref{thm:main} then tells us that this forces the entanglement wedge of the radiation system to include a portion of the black hole interior. We illustrate this in figure \ref{fig:Islandappearance}.\footnote{While in general this implication runs in only one direction, for the constant tension solutions of section \ref{sec:constantT} we find that the island forms if and only if the black hole interior is causally connected to the radiation system.}

In section \ref{sec:discussion} we conclude with some open questions and remarks. 

\section{Review of AdS/BCFT}\label{sec:background}

The $1\rightarrow 2$ theorem will be proven using the focusing theorem for asymptotically AdS spacetimes which feature an ETW brane. In the context of our quantum information discussion however, and in the context of applying our theorem to islands, we have a particular holographic dual description of these spacetimes in mind. We describe this boundary picture in this section.

A BCFT is a conformal field theory living on a manifold with boundary, along with a conformally invariant boundary condition. For appropriate BCFTs, the AdS/BCFT \cite{takayanagi2011holographic, Fujita2011} correspondence suggests a bulk dual description, which consists of an asymptotically AdS region along with an extension of the CFT boundary into the bulk as an end-of-the-world (ETW) brane. To avoid confusion with the bulk-boundary language of the AdS/CFT correspondence, we will refer to the CFT boundary as the \emph{edge}. 
The bulk spacetime and brane are described by an action
\begin{align}\label{eq:ETWaction}
    I_{\text{bulk}}+I_{\text{brane}} & = \frac{1}{16\pi G_N} \int \D^{d+1}x \,\sqrt{g} (R-2\Lambda+L_{\text{matter}}) \notag \\ & \qquad + \frac{1}{8\pi G_N} \int_\mathcal{B} \D^dy\,\sqrt{h} (K + L_{\text{matter}}^\mathcal{B})\;, 
\end{align}
where $L_{\text{matter}}$ and $L_{\text{matter}}^\mathcal{B}$ are matter Lagrangians for fields in the bulk and brane respectively. As usual, $R$ is the Ricci curvature and $\Lambda$ the bulk cosmological constant, while $K$ is the trace of the extrinsic curvature of the brane,
\begin{equation}
    K_{ab} = \nabla_a n_b\;,
\end{equation}
for outward normal $n_j$ to $\mathcal{B}$, and $a, b$ refer to brane coordinates $y^a$.
This action leads to Einstein's equations in the bulk, along with the boundary condition
\begin{align}\label{eq:branegeneralBC}
     -\frac{1}{8\pi G_N} (K_{ab}-Kh_{ab}) = T_{ab}^\mathcal{B}\;.
\end{align}

In AdS/BCFT, the Ryu-Takayanagi formula \cite{Ryu2006b} and its covariant generalization the HRT formula \cite{hubeny2007covariant} continue to calculate the entropy of boundary subregions, provided the homology condition is appropriately adapted \cite{sully2020bcft}. In the context of AdS/CFT, and assuming the null energy condition, the HRT formula is equivalent to the maximin formula \cite{wall2014maximin}. We will assume this remains the case in AdS/BCFT. The maximin formula states that, to leading order in $1/G_N$,
\begin{align}
    S(A) = \max_{\Sigma} \min_{\gamma_A} \left( \frac{\text{Area}[\gamma_A]}{4G_N} \right).
\end{align}
The maximization is over Cauchy surfaces that include $A$ in their boundary, and the minimization is over spacelike codimension 2 surfaces $\gamma_A$ which are homologous to $A$. We will refer to the surface $\gamma_A$ picked out by such a procedure, whose area computes the entropy, as an \emph{entangling surface}. In spacetimes with an ETW brane we should understand the homology constraint as
\begin{align}
    \partial S = \gamma_A \cup A \cup b
\end{align}
for $S$ a spacelike codimension 1 surface in the bulk, and where $b$ is allowed to be any portion of the ETW brane. For a single interval in the CFT, this allows two qualitatively distinct classes of entangling surface: those which do not include a portion of the brane to satisfy the homology constraint, which we call brane-detached, and those which do, which we call brane-attached (see figure \ref{fig:phasetransition3d}). 

In the BCFT description there are degrees of freedom which live at the edge and are associated with the choice of boundary condition. At least for constant, large tension branes these edge degrees of freedom are dual in the bulk to degrees of freedom living on the brane \cite{randall1999alternative,Rozali:2019day}. 

\section{Quantum tasks argument}\label{sec:QIargument}

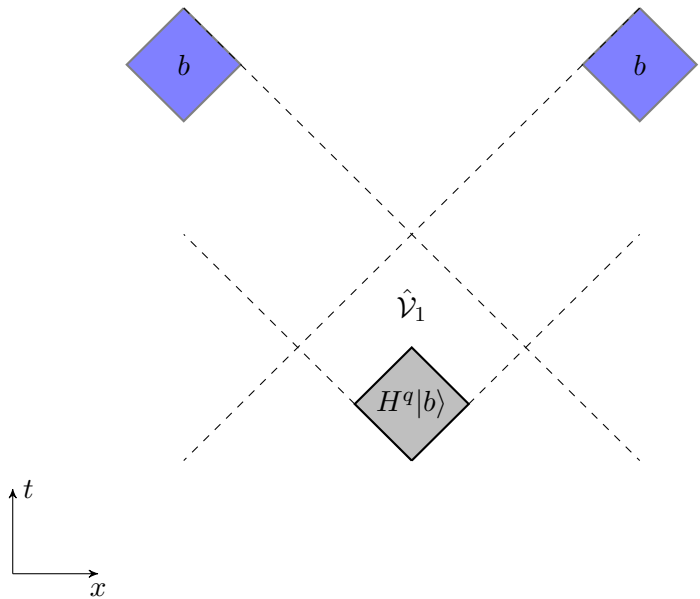
\begin{figure}
    \centering
    \begin{tikzpicture}[scale=0.75]

    \draw[thick,fill=lightgray] (0,0) -- (1,1) -- (0,2) -- (-1,1) -- (0,0);
    \node at (0,1) {$H^q\ket{b}$};

    \node at (0,2.75) {$\hat{\mathcal{V}}_1$};
    
    \draw[fill=blue,opacity=0.5,thick] (4,8) -- (3,7) -- (4,6) -- (5,7) -- (4,8);
    \node at (4,7) {$b$};
    
    \draw[fill=blue,opacity=0.5,thick] (-4,8) -- (-3,7) -- (-4,6) -- (-5,7) -- (-4,8);
    \node at (-4,7) {$b$};
    
    \draw[->] (-7,-2) -- (-7,-0.5);
    \draw[->] (-7,-2) -- (-5.5,-2);
    \node[below] at (-5.5,-2) {$x$};
    \node[right] at (-7,-0.5) {$t$};
    
    \draw[dashed] (0,0) -- (4,4);
    \draw[dashed] (0,0) -- (-4,4);
    
    \draw[dashed] (4,8) -- (-4,0);
    \draw[dashed] (-4,8) -- (4,0);

    \end{tikzpicture}
    \caption{The $\textbf{M}$ task, which we employ to argue for the $1\rightarrow 2$ connected wedge theorem. At $\mathcal{C}_1$ the quantum system $A$ is received which holds a state $H^q\ket{b}$. For the task to be completed successfully, $b$ should be produced at both $r_1$ and $r_2$. We show that completing the task with a high success probability requires the bit $q$ be available in the region $\hat{\mathcal{V}}_1 = \hat{J}^+(\hat{\mathcal{C}}_1)\cap \hat{J}^-(\hat{\mathcal{R}}_1)\cap \hat{J}^-(\hat{\mathcal{R}}_2)$.}
    \label{fig:MTask}
\end{figure}

In this section we give the quantum tasks argument for Theorem \ref{thm:main}. Several aspects of the argument follow the argument of the $2\rightarrow 2$ connected wedge theorem \cite{may2019holographic,may2021holographic}, but we emphasize that the qualitative picture of how the boundary completes the task is distinct in the two cases. In particular, in the $2\rightarrow 2$ theorem the boundary uses a quantum non-local computation to complete the task, whereas in the $1\rightarrow 2$ theorem the boundary employs bulk reconstruction, as we will see below. 

\subsection{The monogamy task}

A quantum task is a quantum computation which has its inputs and outputs at specified spacetime locations. We will consider tasks which have inputs and outputs recorded into extended spacetime regions. To understand this more precisely, we will say that a quantum system $A$ is \emph{localized to region $X$ relative to a channel $\mathcal{M}_X$} if acting on $X$ with $\mathcal{M}_X$ produces $A$. If there exists a channel such that quantum system is localized to a region $X$ relative to that channel, we say just that $A$ is localized to $X$. If it is not possible to learn anything about $A$ from $X$, we say $A$ is \emph{excluded} from $X$. For a review of quantum tasks as they are employed here, see \cite{may2021holographic}.

For our particular example, there is one input region $\hat{\mathcal{C}}_1$ and two output regions $\hat{\mathcal{R}}_1,\hat{\mathcal{R}}_2$. System $A$ is in one of the states $H^q\ket{b}_A$ and is localized to region $\hat{\mathcal{C}}_1$. $H$ is the Hadamard operator, and $b,q\in\{0,1\}$. There is an additional system $Q$ which holds the bit $q$, and we leave unspecified for the moment where $Q$ is located in spacetime. To complete the task the bit $b$ should be localized to $\hat{\mathcal{R}}_1$ and $\hat{\mathcal{R}}_2$. We will momentarily leave the channels $\mathcal{M}_{\hat{\mathcal{C}}_1},\mathcal{M}_{\hat{\mathcal{R}}_1},\mathcal{M}_{\hat{\mathcal{R}}_2}$ unspecified. This task is illustrated in figure \ref{fig:MTask}, and we refer to it as the $\mathbf{M}$ task or ``monogamy task'', for reasons that will become apparent.

We will need to introduce an equivalent formulation of $\mathbf{M}$ that we refer to as \emph{purified} $\mathbf{M}$. The purified task is modified in two ways: (1) a second system $\bar{Q}$ is introduced, and placed in the maximally entangled state with $Q$; and (2) the input qubit $H^q\ket{b}_A$ is replaced with the $A$ system of a maximally entangled state $\ket{\Psi^+}_{A\bar{A}}$. We refer to the $\bar{Q}\bar{A}$ system as the reference system. Notice that Bob can now perform measurements on the reference system to return this to the original task. To do this, Bob first measures the $\bar{Q}$ system, and obtains some output $q$. Then, he measures $\bar{A}$ in the computational basis if $q=0$, and in the Hadamard basis if $q=1$. Bob obtains one bit $b$ of output. Meanwhile, the post-measurement state on $QA$ is $\ket{q}_Q\otimes H^q\ket{b}_A$, so that the inputs are as in the unpurified task. Alice's success probability is unaffected whether Bob performs these measurements before or after Alice returns her outputs, since the $QA$ and $\bar{Q}\bar{A}$ systems never interact. Thus, the purified and unpurified tasks have the same success probability. 

The three regions $\hat{\mathcal{C}}_1,\hat{\mathcal{R}}_1,\hat{\mathcal{R}}_2$ have a naturally associated spacetime region which we label $\hat{\mathcal{V}}_1$, defined according to
\begin{align}
    \hat{\mathcal{V}}_1 \equiv \hat{J}^+(\hat{\mathcal{C}}_1) \cap \hat{J}^-(\hat{\mathcal{R}}_1) \cap \hat{J}^-(\hat{\mathcal{R}}_2)\;.
\end{align}
and which we call the \emph{decision region}. $\hat{\mathcal{V}}_1$ is natural to consider because it is where it is possible to act on $A$ and reach both of $\hat{\mathcal{R}}_1$ and $\hat{\mathcal{R}}_2$. We will in particular be interested in two situations: (1) the setting where $Q$ is localized to $\hat{\mathcal{V}}_1$ and (2) the setting where $Q$ is excluded from $\hat{\mathcal{V}}_1$. 

Let us consider first the case where $Q$ is localized to $\hat{\mathcal{V}}_1$. For convenience, take the unpurified task. Then within $\hat{\mathcal{V}}_1$ Alice should apply $H^q$ to $A$ to obtain $(H^q)^2\ket{b}_A=\ket{b}_A$, measure $\ket{b}_A$ in the $\{\ket{0},\ket{1}\}$ basis, and then send the outcome to each of $r_1$ and $r_2$. Doing so, she can complete the task with high probability, say $p_{\text{suc}}(\mathbf{M})=1-\epsilon$. We introduce the parameter $\epsilon$ to account for the effect of any noise present in carrying out this protocol.\footnote{One source of noise may be our assumption that Alice is working in a classical geometry. In the AdS/CFT context, at finite $G_N$, it seems plausible that small errors are inevitable.}

We can make a stronger statement by introducing a parallel repetition of the monogamy task, which we call $\mathbf{M}^{\times n}$. We consider $n$ states $\{H^{q_i}\ket{b_i}\}_i$ being input at $\hat{\mathcal{C}}_1$, with the $q_i$ and $b_i$ drawn independently and at random. To complete the task, a fraction $1-\delta$ of the $b_i$ should be localized to both $\hat{\mathcal{R}}_i$. As discussed in the last paragraph, Alice can complete each of the $n$ runs with a probability $p_{suc}(\mathbf{M})=1-\epsilon$. For $\epsilon < \delta$, the probability that this leads to more than a fraction $1-\delta$ of the runs being successful will be high. For concreteness take $\delta = 2\epsilon$. In this case we have, at large $n$, 
\begin{align}
    p_{suc}(\mathbf{M}^{\times n}) = 1 - 2 \epsilon^{2+n}.
\end{align}
In particular we see that the success probability converges to $1$ exponentially in $n$.

Next, consider the case where $q$ is excluded from $\hat{\mathcal{V}}_1$. More precisely, we consider purified $\mathbf{M}$ and state this assumption as
\begin{align}
    I(\hat{\mathcal{V}}_1:\bar{Q})=0\;.
\end{align}
Then Alice will be limited in her ability to complete the task, a fact we formalize in the following lemma. 
\begin{lemma}\label{lemma:sucbound}
Consider the $\mathbf{M}$ task [cf. figure \ref{fig:MTask}] with $I(\hat{\mathcal{V}}_1:\bar{Q})=0$. Then any strategy for completing the task has $p_{suc}(\mathbf{M})\leq \cos^2 (\pi/8)$.
\end{lemma}
To see why this is true, consider that Alice holds the $A$ subsystem of a maximally entangled state on $A\bar{A}$ in the region $\hat{\mathcal{V}}_1$. After applying a quantum channel to $A$, she will send part of the output, call it $B_1$, to $\hat{\mathcal{R}}_1$ and part of the output, call it $B_2$, to $\hat{\mathcal{R}}_2$. At best, Alice will learn $Q$ in the regions $\hat{\mathcal{R}}_i$. At each of the $R_i$ then she can use $B_i$ along with $q$ to produce a guess for $b$. This is exactly the guessing game analyzed in \cite{tomamichel2013monogamy}, known as the monogamy of entanglement game. The stated bound on success probability was proven there. 

Notice that if $B_1$ and $\bar{A}$ are maximally entangled, Alice can measure in the $q$ basis and produce an output at $\hat{\mathcal{R}}_1$ which is perfectly correlated with Bobs measurement outcome. Similarly if $B_2\bar{A}$ is maximally entangled she can produce the correct output at $\hat{\mathcal{R}}_2$. The monogamy of entanglement however ensures that there will be a trade-off, and no perfect strategy will exist.\footnote{This explains our naming convention $\mathbf{M}$ for the task.} The proof in \cite{tomamichel2013monogamy} makes this rigorous. 

We can also consider the parallel repetition of the task $\mathbf{M}^{\times n}$ in the case where $I(V:\bar{Q})=0$. Following the reasoning of Lemma \ref{lemma:sucbound}, this can again be reduced to the guessing game discussed in \cite{tomamichel2013monogamy}, who proved that this parallel repetition of the task satisfies the following lemma.  \begin{lemma}\label{lemma:sucboundparallel}
Consider the $\mathbf{M}^{\times n}$ task with $I(\hat{\mathcal{V}}_1:\bar{Q})=0$, and require that a fraction $1-\delta$ of the individual $\mathbf{M}$ tasks are successful. Then any strategy for completing the task has 
\begin{align}
    p_{suc}\leq \left(2^{h(\delta)}\cos^2\left( \frac{\pi}{8}\right) \right)^n \equiv \left(2^{h(\delta)}\beta\right)^n
\end{align}
where $h(\delta)$ is the binary entropy function $h(\delta)\equiv -\delta\log \delta - (1-\delta)\log(1-\delta)$ and the second equality defines $\beta$.
\end{lemma}
For small enough $\delta$ we have that $2^{h(\delta)}\beta<1$, so that with zero mutual information the success probability is small. Our next result will be to show that a large success probability implies a large mutual information. 

In fact, this argument was already completed in \cite{may2021holographic}, albeit in a changed setting.\footnote{In particular the systems $\hat{\mathcal{V}}_1$ and $\bar{Q}$ play the role of systems $\hat{\mathcal{V}}_1$ and $\hat{\mathcal{V}}_2$ discussed in \cite{may2021holographic}. Our Lemma \ref{lemma:mutualinfobound} is their Lemma 7 with this replacement made.}
\begin{lemma}\label{lemma:mutualinfobound}
Suppose that the $\mathbf{M}^{\times n}$ task is completed with success probability $p_{suc}=1-2\epsilon^{2+n}$, where we deem the $\mathbf{M}^{\times n}$ task successful if a fraction $1-2\epsilon$ of the individual $\mathbf{M}$ tasks are. Then the bound
\begin{align}
    \frac{1}{2}I(\hat{\mathcal{V}}_1:\bar{Q})\geq n(-\log 2^{h(2\epsilon)}\beta) - 1 +O((\epsilon/\beta)^n)
\end{align}
holds.
\end{lemma}
This will be the key technical result in the argument from quantum tasks for the connected wedge theorem, which we present in the next section.

We should highlight an important assumption made in proving Lemma \ref{lemma:mutualinfobound}. In addition to the region $\hat{\mathcal{V}}_1$, there is also the spacelike complement $X = [J^+(\hat{\mathcal{V}}_1)\cup J^-(\hat{\mathcal{V}}_1)]^c$. Lemma \ref{lemma:sucboundparallel}, on which Lemma \ref{lemma:mutualinfobound} relies, assumes that information from this region is not made use of in Alice's protocol. If it were, one could use protocols of the type considered in appendix B of \cite{may2019holographic} to perform the $\mathbf{M}^{\times n}$ task without entanglement between $\hat{\mathcal{V}}_1$ and $\bar{Q}$. As discussed in \cite{may2019holographic}, it seems sensible to assume such strategies are not allowed. In particular they require large amounts of GHZ type entanglement in the CFT, which is not expected to exist \cite{nezami2020multipartite}.  

\subsection{Tasks argument for the \texorpdfstring{$1\rightarrow 2$}{TEXT} connected wedge theorem}

With Lemma \ref{lemma:mutualinfobound} in hand, we are ready to complete the tasks argument for the $1\to 2$ connected wedge theorem. For convenience we repeat the theorem here. 

\vspace{0.1cm}
\noindent \emph{\textbf{Theorem \ref{thm:main}:}
Consider three boundary regions $\hat{\mathcal{C}}_1,\hat{\mathcal{R}}_1,\hat{\mathcal{R}}_2$ in an asymptotically AdS$_{2+1}$ spacetime with an end-of-the-world brane. Require that $\hat{\mathcal{C}}_1\subseteq \hat{\mathcal{V}}_1$, and that $\hat{\mathcal{R}}_1,\hat{\mathcal{R}}_2$ touch the brane. Then if $J_{1\rightarrow 12}^{\mathcal{E}}$ is non-empty, the entanglement wedge of $\hat{\mathcal{V}}_1$ is attached to the brane. 
}
\vspace{0.1cm}

\begin{argument}
Using our assumption that $J^\mathcal{E}_{1\rightarrow 12}\neq \varnothing$, we have that there exist bulk points $c_1,r_1,r_2$ such that
\begin{align}
    J^+(c_1) \cap J^-(r_1) \cap J^-(r_2) \cap \mathcal{B} \neq \varnothing
\end{align}
with $c_1\in \mathcal{C}_1$, $r_1\in \mathcal{R}_1,r_2\in \mathcal{R}_2$, where recall $X = \mathcal{E}_W(\hat{X})$, with $\hat{X}$ a boundary region. We will consider a $\mathbf{M}^{\times n}$ task in the bulk such that the input system $A=A_1....A_n$ is input near $c_1$, and each bit $b_i$ should be brought near $r_1$ and $r_2$. Further, system $Q$ will be recorded into the brane degrees of freedom.

It is easy to see that the $\mathbf{M}^{\times n}$ task can be completed in this case with high probability. To see this, note that a simple bulk strategy is to bring $A$ to the brane, learn the $q_i$, and use them to recover the $b_i$. The $b_i$ are then copied and sent to both $r_1$ and $r_2$. Doing so we can complete each $\mathbf{M}$ task with some probability $p=1-\epsilon$, leading to a success probability $p_{suc}=1-2\epsilon^{2+n}$ for the $\mathbf{M}^{\times n}$ task. Since the boundary reproduces bulk physics, the boundary must also complete the task with the same probability. Lemma \ref{lemma:mutualinfobound} then gives 
\begin{align}\label{eq:repeatedbound}
    \frac{1}{2}I(\hat{\mathcal{V}}_1:\bar{Q})\geq n(-\log 2^{h(2\epsilon)}\beta) - 1 +O((\epsilon/\beta)^n)
\end{align}
so that when the entanglement scattering region is non-empty, we have large mutual information. 

This bound on mutual information actually requires the entanglement wedge of $\hat{\mathcal{V}}_1$ to attach to the brane. To see this, consider that in the purified $\mathbf{M}^{\times n}$ task there are $n$ Bell pairs $\ket{\Psi^+}_{A_i\bar{A}_i}$ with $A=A_1...A_n$ input at $\mathcal{C}_1$, and $\bar{A}=\bar{A}_1...\bar{A}_n$ held by Bob. There are an additional $n$ Bell pairs $\ket{\Psi^+}_{Q_i\bar{Q}_i}$, with $Q=Q_1...Q_n$ stored on the brane, and $\bar{Q}=\bar{Q}_1...\bar{Q}_n$ held by Bob. We can choose $n$ to satisfy $O(1)<n<O(1/G_N)$, so that $n$ grows as $G_N\rightarrow 0$ but does so more slowly than $1/G_N$.

Suppose that $\mathcal{E}_W(\hat{\mathcal{V}}_1)$ is not connected to the brane. Then the entropies of the region $\hat{\mathcal{V}}_1$ and of system $\bar{Q}$ satisfy
\begin{align}
    S(\hat{\mathcal{V}}_1) &= \frac{A_{dis}}{4G_N} + n + O(1), \nonumber \\
    S(\bar{Q}) &= n, \nonumber \\
    S(\hat{\mathcal{V}}_1\bar{Q}) &= \frac{A_{dis}}{4G_N} + 2n + O(1). 
\end{align}
The first statement is just our assumption: the disconnected surface calculates the entropy of $\hat{\mathcal{V}}_1$, and then we add the entropy of the $n$ Bell pairs shared between $\hat{\mathcal{V}}_1$ and $\bar{A}$, along with any $O(1)$ contribution. The second statement is due to $Q\bar{Q}$ being in the maximally entangled state. The third statement follows from the disconnected surface being of minimal area along with our choice to take $n<O(1/G_N)$. This is because the other option, of having the connected surface calculate the entropy, would imply that the quantum extremal surface has moved to enclose the $n$ qubits of $Q$, which would happen only if $n > (A_{dis} - A_{con})/G_N$. Using these statements about the entropy, the mutual information is
\begin{align}
    I(\bar{Q}:\hat{\mathcal{V}}_1) &= S(\bar{Q}) + S(\hat{\mathcal{V}}_1) - S(\hat{\mathcal{V}}_1\bar{Q}) = O(1),
\end{align}
so that in the disconnected phase the mutual information is $O(1)$. Since \ref{eq:repeatedbound} implies the mutual information is $O(n)>O(1)$, we find that the entanglement wedge must attach to the brane. 
\end{argument}

It is interesting to consider this result in the context of entanglement wedge reconstruction. We can observe that when the entanglement wedge connects to the brane $Q$ is reconstructable from $\hat{\mathcal{V}}_1$. This clarifies how the boundary completes the task. Whenever the task can be completed in the bulk, the entanglement wedge connects to the brane, which means $Q$ is available in $\hat{\mathcal{V}}_1$. Thus the boundary dynamics can recover the bits $q_i$ and use them to decode the $b_i$, then forward the $b_i$ to both output points.

We should contrast this boundary picture with the analogous feature of the connected wedge theorem in AdS/CFT. In that setting there are two decision regions $\hat{\mathcal{V}}_1$ and $\hat{\mathcal{V}}_2$, with $\hat{\mathcal{V}}_1$ associated with the input $H^q\ket{b}$ and $\hat{\mathcal{V}}_2$ associated with the input $q$. In that case, even in the connected phase, $\hat{\mathcal{V}}_1$ does not reconstruct $q$. To complete the task then the boundary must make use of a different strategy. Indeed in \cite{may2019holographic} the authors argued that the boundary dynamics should be understood as a quantum non-local computation. 

\section{Proof from the focusing theorem}\label{sec:GRproof}

In this section we prove the $1\to 2$ connected wedge theorem. Following \cite{may2019holographic,may2021holographic} closely, our main tools are the focusing theorem and the maximin statement of the HRRT formula \cite{wall2014maximin}. We apply the focusing theorem to null congruences beginning on extremal surfaces. Doing so, new complications arise from the presence of the ETW brane. In particular additional boundary terms arise where the congruence meets the brane. In the next section, we review the usual statement of the focusing theorem, then treat these additional boundary terms.

\subsection{The focusing theorem with boundaries}

We will briefly review the focusing theorem without an ETW brane present. 

Consider a null codimension 1 surface $\mathcal{N}$. We assume $\mathcal{N}$ is foliated by null geodesics which start on a spacelike codimension two surface $\Sigma_1$, and end on another spacelike codimension two surface $\Sigma_2$. Call the affine parameter along the null geodesics $\lambda$, which we scale so that $\lambda=0$ on $\Sigma_1$ and $\lambda=1$ on $\Sigma_2$. Then
\begin{align}
    A(\Sigma_2) - A(\Sigma_1) = \int dY  \sqrt{h}_{\lambda=0}  - \int dY \sqrt{h}_{\lambda=1} = \int_{0}^1 d\lambda \int dY \partial_\lambda \sqrt{h}
\end{align}
where $h$ is the determinant of the induced metric on a constant $\lambda$ slice of $\mathcal{N}$, and $dY = dy^1\wedge ...\wedge dy^{d-2}$. 

Define the expansion, $\theta$, and a $d-1$ form $\bm{\epsilon}$ by
\begin{align}
    \theta = \frac{1}{\sqrt{h}} \partial_\lambda \sqrt{h}\,\,\,\,\,\,\,\,\,\,\,\,\,\,\, \bm{\epsilon} = \sqrt{h} \,d\lambda \wedge dY.
\end{align}
Then the area difference can be written as
\begin{align}\label{eq:naiveStokes}
    A(\Sigma_2) - A(\Sigma_1) = \int \bm{\epsilon} \, \theta.
\end{align}
Expressing the area difference in this way is convenient, since the expansion is constrained in certain situations if we assume the null energy condition (NEC),
\begin{align}
    k^\mu k^\nu T_{\mu\nu} \geq 0.
\end{align}
In particular, consider an extremal surface $\gamma$. Then the boundary of the future or past of $\gamma$, $\partial J^\pm(\gamma)$, is generated by a congruence of null geodesics. Assuming the NEC, this congruence has non-positive expansion when moving away from $\gamma$, as can be shown using the Raychaudhuri equation. We will call surfaces with non-positive expansion \emph{light sheets}. Considering $\mathcal{N}$ to be a portion of either $\partial J^{+}(\gamma)$ or $\partial J^{+}(\gamma)$ then allows us to conclude $A(\Sigma_2)\leq A(\Sigma_1)$. That is, the area of a cross section of the congruence decreases as we follow the geodesics. 

Notice that there are various lightsheets we can define given an extremal surface $\gamma$, in particular the four surfaces $\partial J^\pm_{in,out}(\gamma)$. To specify it will be more convenient to define light sheets as the boundary of the future or past of a (codimension 0) entanglement wedge, $\partial J^{\pm}(X)$. Notice that $\partial J^\pm_{in}(\gamma_X) = \partial J^{\pm}(X)$, so that defining light sheets in this way chooses the inward pointing sheets. 

\begin{figure}
    \centering
    \subfloat[\label{fig:intobrane}]{
    \begin{tikzpicture}[scale=1]
    
    \draw[thick] (0,3) -- (5,3);
    \draw[thick] (2,0) -- (5,0);
    \draw[thick,gray] (0,3) -- (2,0);
    
    \draw[-triangle 45,blue] (1.5,0.75) -- (1.5,0);
    \node[below] at (1.5,0) {$\partial_\lambda$};
    
    \draw[-triangle 45,gray] (1.5,0.75) -- (1,0.25);
    \node[left] at (1,0.25) {$\hat{n}$};
    
    \foreach \x in {0,...,15}
    {
    \draw[blue,opacity=0.4,postaction={on each segment={mid arrow}}] ({2+\x/5},3) -- ({2+\x/5},0);
    }
    
    \foreach \x in {0,...,9}
    {
    \draw[blue,opacity=0.4,postaction={on each segment={mid arrow}}] ({\x/5},3) -- ({\x/5},{3-\x/3.333});
    }
    
    \node[right] at (5,3) {$\Sigma_1$};
    \node[right] at (5,0) {$\Sigma_2$};
    
    \node[right] at (0.5,1.25) {$\mathcal{B}$};
    
    \end{tikzpicture}
    }
    \hfill
    \subfloat[\label{fig:outofbrane}]{
    \begin{tikzpicture}[scale=1]
    
    \draw[thick] (2,3) -- (5,3);
    \draw[thick] (0,0) -- (5,0);
    \draw[thick,gray] (2,3) -- (0,0);
    
    \foreach \x in {0,...,15}
    {
    \draw[blue,opacity=0.4,postaction={on each segment={mid arrow}}] ({2+\x/5},3) -- ({2+\x/5},0);
    }
    
    \foreach \x in {0,...,9}
    {
    \draw[blue,opacity=0.4,postaction={on each segment={mid arrow}}] ({\x/5},{\x/3.333}) -- ({\x/5},{0});
    }
    
    \node[right] at (5,3) {$\Sigma_1$};
    \node[right] at (5,0) {$\Sigma_2$};
    
    \node[right] at (0.4,1.5) {$\mathcal{B}$};
    
    \draw[blue,-triangle 45] (1.5,2.25) -- (1.5,1.5);
    \node[above right] at (1.5,1.5) {$\partial_\lambda$};
    
    \draw[gray,-triangle 45] (1.5,2.25) -- (1,2.75);
    \node[left] at (1,2.75) {$\hat{n}$};
    
    \end{tikzpicture}
    }
    \caption{A portion of the boundary of the past $\partial J^-(\mathcal{R}_i)$, showing two cross sections $\Sigma_1$, $\Sigma_2$, and the end-of-the-world brane $\mathcal{B}$. Null geodesics generating the lightcone are shown in blue. The outward pointing normal to the brane is labelled $\hat{n}$, while the tangent vector to the null geodesics is labelled $\partial_\lambda$. (a) When $\hat{n}\cdot \partial_\lambda \geq 0$, the brane removes area. (b) When $\hat{n}\cdot \partial_\lambda \leq 0$, the brane adds area.}
    \label{fig:boundaryterms}
\end{figure}
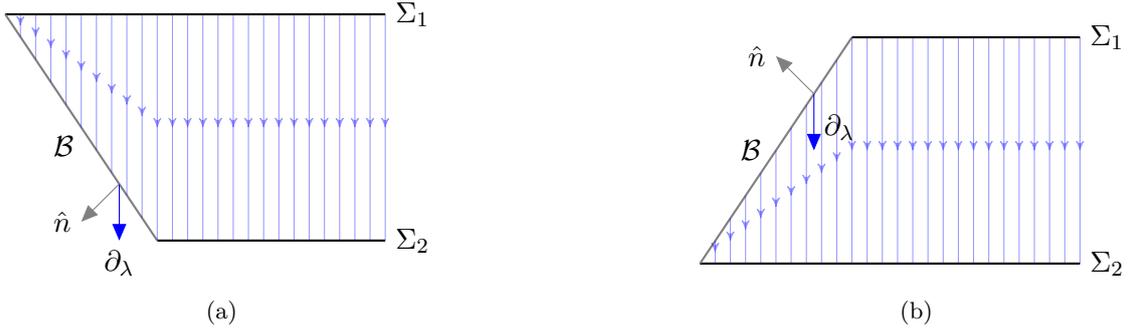

Next, we consider the focusing theorem in the setting where $\mathcal{N}$ intersects the brane. The situation is shown in figure \ref{fig:boundaryterms}. The null surface $\mathcal{N}$ is still foliated by a null congruence, but some geodesics end or begin on an additional portion of the boundary, $\mathcal{N}\cap \mathcal{B}$. To prove an area theorem in this setting, we will need to assume the NEC holds both for the bulk stress tensor and for the branes stress tensor. This later statement is
\begin{align}
    \ell^a \ell^b T_{ab}^\mathcal{B} \geq 0,
\end{align}
where $\ell^a$ is a null tangent vector to the brane. This is satisfied with equality for constant tension branes. Using the boundary condition $8\pi G_NT_{ab}= -(K_{ab}-Kh_{ab})$ we can also express this as $\ell^a \ell^b K_{ab}\leq 0$.

We will reconsider $\int \bm{\epsilon}\, \theta$ and write this as a boundary integral. A simple application of the fundamental theorem of calculus sufficed to derive \ref{eq:naiveStokes}, but this was only because the null geodesics meet $\Sigma_1$ and $\Sigma_2$ normally. For the additional portion of the boundary we need to use Stokes theorem in a more general form. To begin, note that
\begin{align}
    \epsilon \,\theta = (\partial_\lambda \sqrt{h}) \, d\lambda \wedge dY = d(\sqrt{h} dY) \equiv d\omega
\end{align}
so $\bm{\epsilon} \,\theta$ is closed. The last equality defines $\omega$. Now we will use Stokes theorem in the form
\begin{align}\label{eq:stokes}
    \int_M d\omega = \int_{\partial M} d^{d-2}x \sqrt{\gamma} n^\mu V_\mu
\end{align}
where $\gamma$ is the induced metric on the boundary, $n^\mu$ is the normal vector to the boundary\footnote{For spacelike boundaries we should choose the outward pointing normal, while for timelike boundaries we choose the inward pointing one.}, and $V_\mu = (-1)^{d-1}(*\omega)_\mu$ where $*$ denotes the Hodge dual. 

The one-form $\bm{V}$ in \ref{eq:stokes} is simple to compute, $\bm{V} = (-1)^{d-1}*\omega = (-1)^{d-1}d\lambda$. Along $\Sigma_1$ we have $n^\mu = -(\partial_\lambda)^\mu$, and along $\Sigma_2$ we have $n^\mu=(\partial_\lambda)^\mu$, which recovers the two boundary terms appearing in \ref{eq:naiveStokes}. The boundary $\mathcal{B} \cap \mathcal{N}$ returns an additional term, 
\begin{align}\label{eq:stokesexpanded}
    \int_{\mathcal{N}}\bm{\epsilon}\, \theta  = A(\Sigma_2) - A(\Sigma_1) + \int_{\Sigma \cap \mathcal{N}} d^{d-2}x \sqrt{\gamma} \,n_\lambda .
\end{align}
We will need this more general statement when we focus backwards in the proof of Theorem \ref{thm:main}.

For \ref{eq:stokesexpanded} to relate $A(\Sigma_2)$ and $A(\Sigma_1)$ we would like to fix the sign of $n_\lambda$. In particular, $n_\lambda \geq 0$ along with $\theta \leq 0$ would imply $A(\Sigma_2) \geq A(\Sigma_1)$, recovering the usual area theorem. This is illustrated in figure \ref{fig:boundaryterms}. In fact we can show $n_\lambda \geq0$ in one particular but important situation. Suppose that $\mathcal{N}$ is a portion of $\partial J^-(\mathcal{R}_i)$, for $\mathcal{R}_i$ the entanglement wedge of an edge anchored region. Then we have that at $\lambda=0$,\footnote{More generally we need only that $n_\lambda\geq 0$. Thus the theorem may still hold in certain cases where $\hat{\mathcal{R}}_i$ is not attached to the edge, though its unclear when this occurs.} 
\begin{align}
    n_\lambda =0.
\end{align}
This holds because the entangling surface $\gamma_{\mathcal{R}_i}$ meets the brane normally, which means the normal vectors of $\gamma_{\mathcal{R}_i}\cap \mathcal{B}$ will be tangent to the brane. 

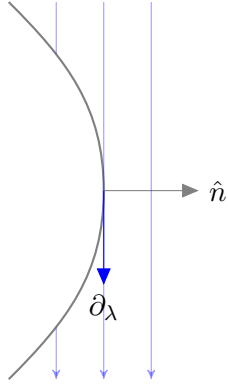
\begin{figure}
    \centering
    \begin{tikzpicture}[scale=1.25]
    
    \draw[gray,thick] (0,0) to [out=-90,in=45] (-1,-2);
    \draw[gray,thick] (0,0) to [out=90,in=-45] (-1,2);
    
    \draw[gray,-triangle 45] (0,0) -- (1,0);
    \node[right] at (1,0) {$\hat{n}$};
    
    \draw[opacity=0.4,blue,->] (0,2) -- (0,-2);
    \draw[opacity=0.4,blue,->] (0.5,2) -- (0.5,-2);
    \draw[opacity=0.4,blue] (-0.5,2) -- (-0.5,1.45); 
    \draw[opacity=0.4,blue,<-] (-0.5,-2) -- (-0.5,-1.45); 
    
    \draw[blue,-triangle 45] (0,0) -- (0,-1);
    \node[below] at (0,-1) {$\partial_\lambda$};
    
    \end{tikzpicture}
    \caption{The lightsheet $\partial J^-(R_i)$ where it meets the brane. For $\hat{n}\cdot \partial_\lambda =n_\lambda \geq 0$ initially, a change in sign requires that the extrinsic curvature be positive somewhere along the brane (as shown here), which is ruled out by the NEC applied to the brane stress tensor.}
    \label{fig:pictureproof}
\end{figure}

We claim the NEC imposed on the brane stress tensor ensures $n_\lambda\geq 0$ everywhere. To see this, study the derivative of $n_\lambda$ as we move along the brane,
\begin{align}
    \ell^\mu\nabla_\mu (n_\lambda) =\ell^\mu \nabla_\mu (n_\sigma k^\sigma) = \ell^\mu k^\sigma \nabla_\mu n_\sigma + \ell^\mu n_\sigma \nabla_\mu k^\sigma
\end{align}
Using
\begin{align}
    k^\sigma &= n_\lambda n^\sigma + \ell^\sigma, \nonumber \\
    0 &= n^\nu \nabla_\mu n_\nu, \nonumber \\
    0 &= k^\mu \nabla_\mu k^\nu ,
\end{align}
this becomes
\begin{align}
    \ell^\mu\nabla_\mu (n_\lambda) = \ell^\mu \ell^\sigma \nabla_\mu n_\sigma - n_\lambda n^\mu n_\sigma \nabla_\mu k^\sigma.
\end{align}
Since initially $n_\lambda=0$, if we establish that $\nabla_\lambda n_\lambda \geq 0$ whenever $n_\lambda=0$, we are done. But when $n_\lambda=0$ the above is just
\begin{align}
    \nabla_\lambda n_\lambda = \ell^\mu \ell^\sigma \nabla_\mu n_\sigma = -\ell^a \ell^b K_{ab} \geq 0
\end{align}
where the minus sign in the second equality is introduced because $K_{ab}$ is defined using the inward pointing normal vector, whereas the normal vector appearing in Stokes theorem was outward pointing. The inequality is just the NEC imposed on the brane. How the curvature in the brane prevents a sign change in $n_\lambda$ is illustrated in figure \ref{fig:pictureproof}. 

\subsection{Proof of the connected wedge theorem}

In this section we prove the $1\rightarrow 2$ connected wedge theorem for asymptotically AdS spacetimes with an ETW brane. Our proof follows the earlier proof for the $2\rightarrow 2$ connected wedge theorem appearing in \cite{may2019holographic,may2021holographic} closely, since a minor modification of the proof given there suffices to prove our theorem. We repeat the full proof in order to explain this modification clearly, and to keep the paper self contained.

The proof relies on three assumptions: (i) that the null energy condition holds in the bulk; (ii) that the null energy condition holds for the branes stress tensor; and (iii) that the maximin procedure \cite{wall2014maximin,marolf2019restricted,akers2020quantum} for finding HRRT surfaces is correct even in the context of AdS/BCFT.

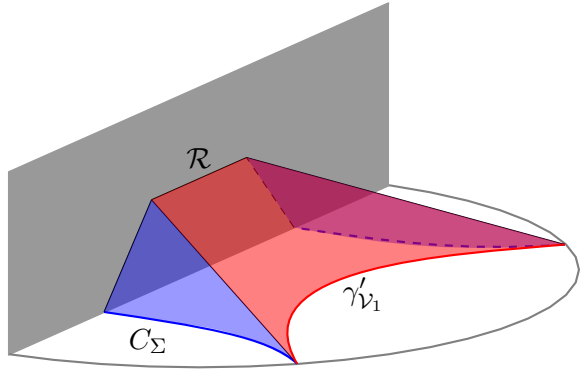
\begin{figure}
    \centering
    \tdplotsetmaincoords{15}{0}
    \begin{tikzpicture}[scale=2.5,tdplot_main_coords]
    \tdplotsetrotatedcoords{0}{30}{0}
    
    \begin{scope}[tdplot_rotated_coords]
    
    \draw[thick,fill=black,opacity=0.4] (0,2,2) -- (0,2,-2) -- (0,3,-2) -- (0,3,2);
    
    \begin{scope}[canvas is xz plane at y=2]
    \draw[domain=-90:90,thick,gray] plot ({2*cos(\x)}, {2*sin(\x)});
    \draw[blue,thick,dashed] (1.41,1.41) to [out=-135,in=0] (0,1);
    \draw[blue,thick] (1.41,-1.41) to [out=135,in=0] (0,-1);
    \end{scope}
    
    
    \draw[black] (1.41,2,1.41) -- (0,2.5,0.5);
    \draw[black] (1.41,2,-1.41) -- (0,2.5,-0.5);
    
    \draw (0,2,-1) -- (0,2.5,-0.5);
    \draw[dashed] (0,2,1) -- (0,2.5,0.5);
    
    \draw (0,2.5,-0.5) -- (0,2.5,0.5);
    
    \fill[blue,opacity=0.4] (1.41,2,-1.41) to [out=146,in=-11] (0,2,-1) -- (0,2.5,-0.5);
    \fill[blue,opacity=0.4] (1.41,2,1.41) to [out=-176,in=-11] (0,2,1) -- (0,2.5,0.5);
    
    \fill[red,opacity=0.5] (1.41,2,1.41) to [out=185,in=120] (1.41,2,-1.41) -- (0,2.5,-0.5) -- (0,2.5,0.5);
    
    \draw[thick,red] (1.41,2,1.41) to [out=185,in=120] (1.41,2,-1.41);
    \node[above] at (0,2.5,0) {$\mathcal{R}$};
    \node at (0.5,2,-1.4) {$C_\Sigma$};
    
    \node at (1,2,0) {$\gamma'_{\mathcal{V}_1}$};
    
    \end{scope}
    \end{tikzpicture}
    \caption{The null membrane. The red surface is the lift $\mathcal{L}$, the blue surfaces make up the slope. The ridge $\mathcal{R}$, is where the lift meets the brane.}
    \label{fig:nullmembrane}
\end{figure}

Given these assumptions, the outline of the proof of Theorem \ref{thm:main} is as follows. We suppose, by way of contradiction, that $J_{1\rightarrow 12}\neq \varnothing$ and the HRRT surface for region $\hat{\mathcal{V}}_1$ is brane-detached. Call this surface $\gamma'_{{\mathcal{V}}_1}$. According to the maximin procedure, this surface is minimal in some Cauchy slice $\Sigma$. We'll use the focusing theorem and the fact that $J_{1\rightarrow 12}^{\mathcal{E}}\neq \varnothing$ to construct a smaller area surface in $\Sigma$ which is brane-connected, called the \emph{contradiction surface} $C_{\Sigma}$. This provides a contradiction with $\gamma'_{{\mathcal{V}}_1}$ having been the HRRT surface, showing the correct HRRT surface must be brane-attached. 

To begin, we consider two cases, corresponding to the boundary scattering region
\begin{align}
    \hat{J}_{1\rightarrow 12}^\mathcal{E} = \hat{J}^+(\hat{\mathcal{C}}_1) \cap \hat{J}^{-}(\hat{\mathcal{R}}_1) \cap \hat{J}^-(\hat{\mathcal{R}}_2)\cap \mathcal{B} = \hat{\mathcal{V}}_{1} \cap \mathcal{B}
\end{align}
being empty or non-empty. If it is non-empty, then $\hat{\mathcal{V}}_1$ is attached to the brane in the boundary, so its entanglement wedge is immediately brane attached and we are done. If it is empty, we proceed with the proof below.

Define the null surface
\begin{align}
    \mathcal{L}= \partial J^+(\mathcal{V}_1) \cap J^-(\mathcal{R}_1)\cap J^-(\mathcal{R}_2)
\end{align}
which we call the \emph{lift}. This is  defined by taking the inward pointing null orthogonal vectors of $\gamma_{\mathcal{V}_1}$ as generators for a null congruence, and extending those geodesics until they reach the past of $\mathcal{R}_1$ or $\mathcal{R}_2$. Additionally, geodesics should not be extended past any caustic points --- defining the lift in terms of $\partial J^+(\mathcal{V}_1)$ implements this for us, as geodesics leave the boundary of $J^+(\mathcal{V}_1)$ after developing a caustic. 

There are two features of the lift that will be important. The first feature is that the lift has a non-empty intersection with the brane. To see this, recall that by assumption
\begin{align}
    J_{1\rightarrow 12}^{\mathcal{E}} = J^+(\mathcal{C}_1) \cap J^-(\mathcal{R}_1)\cap J^-(\mathcal{R}_2) \cap \mathcal{B} \neq \varnothing.
\end{align}
Then, recall that since $\hat{\mathcal{C}}_i\subseteq \hat{\mathcal{V}}_i$, we have also $\mathcal{C}_i\subseteq \mathcal{V}_i$. Thus we learn
\begin{align}
    J_{1\rightarrow 12}^{\mathcal{E}} \subseteq J^+(\mathcal{V}_1) \cap J^-(\mathcal{R}_1)\cap J^-(\mathcal{R}_2) \cap \mathcal{B} \neq \varnothing.
\end{align}
This gives that $J^+(\mathcal{V}_1)$ meets the brane while in the past of $\mathcal{R}_1$ and $\mathcal{R}_2$. In particular then the ridge, defined by
\begin{align}
    \mathcal{R}\equiv \mathcal{L} \cap \mathcal{B} \neq \varnothing
\end{align}
is non-empty. 

The second important feature of the lift is that its boundary has a component $\mathcal{A}_1$ along $\partial J^-(\mathcal{R}_1)$ and a component $\mathcal{A}_2$ along $J^-(\mathcal{R}_2)$ which are separated by the ridge. The other possibility would be for the ridge to extend to one or more of the edges. This cannot occur however, which follows because the ridge is a subregion of the bulk scattering region, which by assumption does not extend to the boundary. 

Next define a second null sheet which we call the \emph{slope},
\begin{align}
    \mathcal{S}_\Sigma = \partial[J^-(\mathcal{R}_1)\cap J^-(\mathcal{R}_2)]\cap J^-[\partial J^+(\mathcal{V}_1)]\cap J^+(\Sigma).
\end{align}
The slope is generated by past-directed null geodesics beginning as the inward, past directed null normals to $\gamma_{{\mathcal{R}_1}}$ and $\gamma_{{\mathcal{R}_2}}$, and extended until they reach $\Sigma$. We will be particularly interested in
\begin{align}
    C_\Sigma \equiv S_{\Sigma}\cap \Sigma
\end{align}
which we introduced above as the contradiction surface. The lift, ridge, slope, and contradiction surface are shown in figure \ref{fig:nullmembrane}.

Now, we apply the focusing theorem in the form of equation \ref{eq:stokesexpanded} to the lift and to the slope. The lift is a portion of the boundary of the future of an extremal surface, $\partial J^+(\mathcal{V}_1)$, so focusing applies. We choose a parameterization such that the null generators begin on $\gamma_{\mathcal{V}_1}$ and end on $\mathcal{R}\cup \mathcal{A}_1\cup \mathcal{A}_2\cup B_{\mathcal{L}}$, where $\mathcal{B}_{\mathcal{L}}$ is any caustics present in the lift. This leads to 
\begin{align}
    \text{area}(\mathcal{A}_2) + \text{area}(\mathcal{A}_1) +2\, \text{area}(B_{\mathcal{L}})+ \text{area}(\mathcal{R}) - \text{area}(\gamma'_{\mathcal{V}_1}) = \int \bm{\epsilon} \, \theta \leq 0.
\end{align}
Similarly, we can apply \ref{eq:stokesexpanded} to the slope, which is a portion of $\partial [J^-(\mathcal{R}_1)\cap J^-(\mathcal{R}_2)]$. Choosing the parameterization such that generators begin on $\mathcal{A}_1\cup \mathcal{A}_2$ and end on $C_\Sigma \cup B_{S_\Sigma}$, where $B_{S_{\Sigma}}$ is any caustics present in the slope. We have then
\begin{align}\label{eq:slopeterms}
    \text{area}(C_\Sigma)+2\,\text{area}(B_{\mathcal{S}_\Sigma}) - \text{area}(\mathcal{A}_2) - \text{area}(\mathcal{A}_1) + \int_{S_\Sigma \cap \mathcal{B}} d^{d-2}x \,\sqrt{\gamma} n_\lambda = \int \bm{\epsilon} \,\theta \leq 0.
\end{align}
Adding these two inequalities and rearranging terms we obtain
\begin{align}
   \text{area}(\gamma'_{\mathcal{V}_1}) &\geq  \text{area}(C_\Sigma) + \text{area}(R) +\text{area}(B_{\mathcal{S}_\Sigma}) +\text{area}(B_{\mathcal{L}}) + \int_{S_\Sigma \cap \mathcal{B}} d^{d-2}x \,\sqrt{\gamma} n_\lambda, \nonumber \\
   & \geq \text{area}(C_\Sigma)
\end{align}
where we've used that $n_\lambda\geq0$, which follows when the NEC applied to the brane matter tensor holds, as shown at the end of the last section. This ensures that the brane-disconnected surface $\gamma'_{\mathcal{V}_1}$ is not of minimal area in the Cauchy slice $\Sigma$, so from the maximin procedure cannot be the correct HRRT surface, completing the proof. 

We should highlight the modifications made from the similar proof of the $2\rightarrow 2$ connected wedge theorem \cite{may2019holographic, may2021holographic}. In that case, there were four regions $\mathcal{C}_1,\mathcal{C}_2,\mathcal{R}_1$ and $\mathcal{R}_2$, and two decision regions $\mathcal{V}_1$ and $\mathcal{V}_2$. The lift was formed by a null congruence of geodesics starting on $\gamma_{\mathcal{V}_1} \cup \gamma_{\mathcal{V}_2}$. Points on the ridge corresponded to where a geodesic starting on $\gamma_{\mathcal{V}_1}$ collided with a geodesic starting on $\gamma_{\mathcal{V}_2}$, whereas in our setting the ridge is formed by generators from $\gamma'_{\mathcal{V}_1}$ colliding with the brane. Another distinction is the occurrence of the boundary $S_{\Sigma}\cap \mathcal{B}$ and associated term in \ref{eq:slopeterms}. This is handled in our case by assuming the NEC holds for the brane stress tensor.  

\subsection{Comments on the \texorpdfstring{$1\rightarrow 2$}{TEXT} connected wedge theorem}

\subsubsection*{The scattering region is inside the entanglement wedge}

In the context of the $2\rightarrow 2$ theorem, \cite{may2019holographic, may2021holographic} showed that the scattering region $J_{12\rightarrow 12}$ is inside of the entanglement wedge of $\hat{\mathcal{V}}_1\cup \hat{\mathcal{V}}_2$. It is straightforward to adapt either of the proofs given there to the $1\rightarrow 2$ theorem, where the analogous statement is that $J_{1\rightarrow 12}^{\mathcal{E}}$ is inside the entanglement wedge of $\hat{\mathcal{V}}_1$. Since $J_{1\rightarrow 12}^{\mathcal{E}}$ lives in the brane, we can be more specific and say that $J_{1\rightarrow 12}^{\mathcal{E}}$ is inside the island formed by $\hat{\mathcal{V}}_1 \cap \mathcal{B}$. 

\subsubsection*{Relationship to $2\rightarrow 2$ theorem and interface branes}

It is possible to describe ETW brane geometries as a $\mathbb{Z}_2$ identification of an interface brane geometry. In particular, consider a spacetime $\mathcal{M}$ described by metric $g_{\mu\nu}(x^\mu)$ and satisfying the boundary condition
\begin{align}
    K_{ab} - K h_{ab} = - 8\pi G_N T_{ab}^\mathcal{B}
\end{align}
at the ETW brane. Then we can define a doubled geometry featuring an interface brane, with metric $g_{\mu\nu}(x^\mu_+)$ on one side of the brane and a copy of that metric $g_{\mu\nu}(x^\mu_-)$ on the other. At the interface brane Einsteins equations require we satisfy the Israel junction conditions
\begin{align}
    h_{ab}^+ &= h_{ab}^-, \\
    [K_{ab}^+ - K_{ab}^-] - [K^+-K^-] h_{ab} &= - 8\pi G_N T^I_{ab}.
\end{align}
Setting $T^I_{ab}=2T_{ab}^B$ satisfies this condition. Identifying points $x_+=x_-$ then recovers the ETW brane geometry. 

We can apply the $2\rightarrow 2$ connected wedge theorem to this interface brane geometry, and in limited cases recover the $1\rightarrow 2$ theorem. To do this choose $\hat{\mathcal{C}}_1$ and $\hat{\mathcal{C}}_2$ to be mirror images across the interface brane. Choose $\hat{\mathcal{R}}_1$ and $\hat{\mathcal{R}}_2$ to be intervals centered on the two CFT interfaces. Notice that the brane anchored scattering region $J_{1\rightarrow 12}^{\mathcal{E}}$ is not empty if and only if the bulk scattering region $J_{12\rightarrow 12}^\mathcal{E}$ in the interface geometry is not empty. Further, the entanglement wedge of $\hat{\mathcal{V}}_1\cup \hat{\mathcal{V}}_2$ will be connected if and only if the entanglement wedge of $\hat{\mathcal{V}}_1$ connects to the brane in the ETW brane geometry. Thus, when the doubled geometry satisfies the conditions for the $2\rightarrow 2$ theorem --- in particular when the NEC holds in the doubled geometry --- the $1\rightarrow 2$ theorem follows from the $2\rightarrow 2$ theorem.

Recall however the conditions for the $1\rightarrow 2$ theorem: the bulk stress tensor and brane stress tensor should separately satisfy the NEC. There are many cases where these conditions hold, but in the associated interface brane geometry the NEC is violated. Consider for instance an ETW brane solution with
\begin{align}
    T_{\mu\nu} &= 0, \nonumber \\
    T^{\mathcal{B}}_{ab} &= -T h_{ab}.
\end{align}
Then in the interface brane geometry the stress tensor is
\begin{align}
    T_I^{\mu\nu} = -T h^{ab} e^{\mu}_a e^\nu_b \,\delta(x-x_0)
\end{align}
where the delta function is turned on at the interface. To study the NEC for $T^I_{\mu\nu}$, it's convenient to rewrite this using the completeness relation,
\begin{align}
    g^{\mu\nu} = n^\mu n^\nu + h^{ab} e^{\mu}_a e^{\nu}_b
\end{align}
so that
\begin{align}
    T_I^{\mu\nu} \ell_\mu \ell_\nu = -T(g^{\mu\nu}-n^\nu n^\mu)\ell_\mu \ell_\nu = T (n^\mu \ell_\mu)^2
\end{align}
We see that the NEC is satisfied if and only if $T>0$. However, in the ETW brane geometry, the $1\rightarrow 2$ theorem holds even for $T<0$. Consequently we find that the $2\rightarrow 2$ theorem applied to the interface geometry only recovers the $1\rightarrow 2$ theorem in special cases. 

\subsubsection*{Counterexample to the converse}

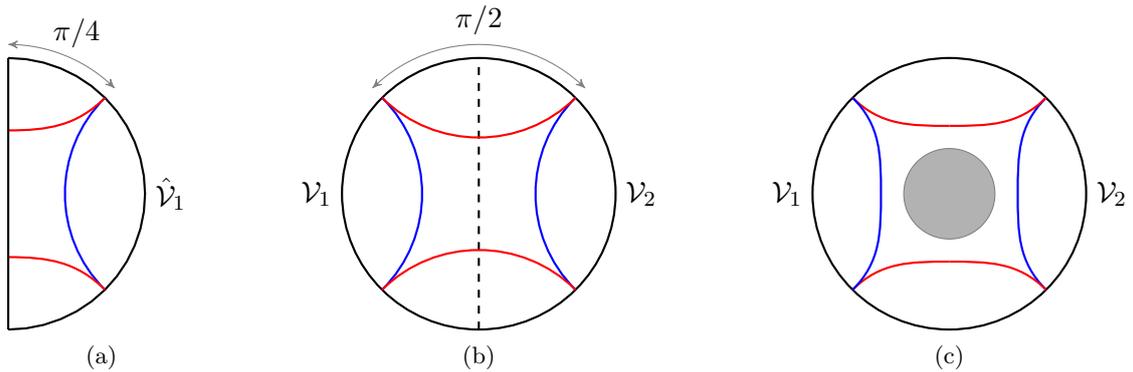
\begin{figure}
    \centering
    \subfloat[\label{fig:halfcircle}]{
    \begin{tikzpicture}[scale=0.6]
    
    \draw[thick,domain=-90:90] plot ({3*cos(\x)},{3*sin(\x)});
    
    \draw[blue, thick] (2.12, 2.12) to [out=-135,in=135] (2.12, -2.12);
    
    \draw[red, thick] (0, 1.4) to [out=0,in=-135] (2.12, 2.12);
    \draw[red, thick] (0, -1.4) to [out=0,in=135] (2.12, -2.12);
    
    \draw[black,thick] (0,-3) -- (0,3);
    
    \node[right] at (3,0) {$\hat{\mathcal{V}}_1$};
    
    \draw[<->,gray,domain=90:45] plot ({3.3*cos(\x)},{3.3*sin(\x)});
    \node[above] at (1.5,3) {$\pi/4$};
    
    \end{tikzpicture}
    }
    \hfill
    \subfloat[\label{fig:circlebalance}]{
    \begin{tikzpicture}[scale=0.6]
    
    \draw[black,thick,dashed] (0,-3) -- (0,3);
    
    \draw[thick] (0,0) circle (3);
    
    \draw[blue, thick] (2.12, 2.12) to [out=-135,in=135] (2.12, -2.12);
    \draw[blue, thick] (-2.12, 2.12) to [out=-45,in=45] (-2.12, -2.12);
    
    \draw[red, thick] (-2.12, 2.12) to [out=-45,in=-135] (2.12, 2.12);
    \draw[red, thick] (-2.12, -2.12) to [out=45,in=135] (2.12, -2.12);
    
    \node[right] at (3,0) {$\mathcal{V}_2$};
    \node[left] at (-3,0) {$\mathcal{V}_1$};
    
    \draw[<->,gray,domain=135:45] plot ({3.3*cos(\x)},{3.3*sin(\x)});
    \node[above] at (0,3.3) {$\pi/2$};
    
    \end{tikzpicture}
    }
    \hfill
    \subfloat[\label{fig:withmatter}]{
    \begin{tikzpicture}[scale=0.6]
    
    \node[right] at (3,0) {$\mathcal{V}_2$};
    \node[left] at (-3,0) {$\mathcal{V}_1$};
    
    \draw[thick] (0,0) circle (3);
    \draw[fill=black!60!,opacity=0.5] (0,0) circle (1);
    
    \draw[blue, thick] (2.12, 2.12) to [out=-135,in=90] (1.5, 0);
    \draw[blue, thick] (1.5, 0) to [out=-90,in=135] (2.12, -2.12);
    
    \draw[red, thick] (2.12, 2.12) to [out=-135,in=0] (0, 1.5);
    \draw[red, thick] (0, 1.5) to [out=-180,in=-45] (-2.12, 2.12);
    
    \draw[red, thick] (2.12, -2.12) to [out=135,in=0] (0, -1.5);
    \draw[red, thick] (0, -1.5) to [out=-180,in=45] (-2.12, -2.12);
    
    \draw[blue, thick] (-2.12, 2.12) to [out=-45,in=90] (-1.5, 0);
    \draw[blue, thick] (-1.5, 0) to [out=-90,in=45] (-2.12, -2.12);
    
    \end{tikzpicture}
    }
    \caption{A counterexample to the converse of Theorem \ref{thm:main}. (a) A constant time slice of a solution with a $T=0$ brane sitting in pure AdS. These solutions are described in detail section \ref{sec:constantT}. We choose a region $\hat{\mathcal{V}}_1$ of size $\pi/2$ and which is centered between the two edges. This region sits exactly on the phase transition between brane-attached (red surface) and brane detached (blue surface), and the scattering region consists of a single point. (b) The $T=0$ solution can be viewed as a $\mathbb{Z}_2$ identification of global AdS with the identification across $\rho=0$. (c) In the unfolded geometry, we consider adding a spherically symmetric matter distribution (shown in grey). This delays light rays travelling from $c_1$ to the brane by some finite amount, closing the scattering region. Due to spherical symmetry, the region $\hat{\mathcal{V}}_1$ remains on the phase transition. Increasing its size infinitesimally then keeps the scattering region closed, while also ensuring the red, brane-attached surface is minimal.}
    \label{fig:counterexample}
\end{figure}

We claimed in the introduction that the converse to Theorem \ref{thm:main} is false. In \cite{may2019holographic, may2021holographic}, the authors constructed a counterexample to the converse of the $2\rightarrow 2$ theorem. By taking a $\mathbb{Z}_2$ identification of the solution used in their example we can easily construct a counterexample to the converse of the $1\rightarrow 2$ theorem. We do this in figure \ref{fig:counterexample}. 

\subsubsection*{The out regions are not entangled}

In the $2\rightarrow 2$ connected wedge theorem, time reversal implies that in addition to the decisions regions having a connected entanglement wedge, an analogous pair of late time regions do as well, where the late time regions are defined by\footnote{We are interested here in the case where $\hat{\mathcal{W}}_i\subseteq \hat{\mathcal{R}}_i$, analogous to our condition $\hat{\mathcal{C}}_i \subseteq \hat{\mathcal{V}}_i$ on the input and decision regions.}
\begin{align}
    \hat{\mathcal{W}}_1 &= J^-(\hat{\mathcal{R}}_1) \cap J^+(\hat{\mathcal{C}}_1) \cap J^+(\hat{\mathcal{C}}_2), \nonumber \\
    \hat{\mathcal{W}}_2 &= J^-(\hat{\mathcal{R}}_2) \cap J^+(\hat{\mathcal{C}}_1) \cap J^+(\hat{\mathcal{C}}_2).
\end{align}
In the context of the $1\rightarrow 2$ theorem one can define similar regions. To do so, we define points $x_1$, $x_2$ as the points where $\partial \hat{J}^+(\hat{\mathcal{C}}_1)$ reaches edge $1$ and edge $2$, respectively. Then we define
\begin{align}
    \hat{\mathcal{W}}'_1 &= \hat{J}^+(x_1) \cap J^-(\hat{\mathcal{R}}_1), \nonumber \\
    \hat{\mathcal{W}}'_2 &= \hat{J}^+(x_2) \cap J^-(\hat{\mathcal{R}}_2).
\end{align}
We can ask if $\hat{\mathcal{W}}'_1$ and $\hat{\mathcal{W}}_2$ must also be entangled when the entanglement scattering region is non-empty. 

In fact, these regions do not need to be entangled. For an explicit counterexample, begin with the example shown in figure \ref{fig:halfcircle}, where $\hat{\mathcal{V}}_1$ consists of an interval of size $\pi/2$ centered between the two edges. Then the scattering region consists of a single point, and the minimal surface enclosing $\hat{\mathcal{W}}'_1\cup \hat{\mathcal{W}}_2'$ is on the transition from giving a connected and disconnected entanglement wedge. Now decrease the tension, moving the brane inward. This shortens the light travel time from $\hat{\mathcal{C}}_1$, so increases the size of the scattering region. Meanwhile, the disconnected surface enclosing $\hat{\mathcal{W}}'_1\cup \hat{\mathcal{W}}_2'$ loses area and becomes dominant, so that there is a non-empty scattering region but only $O(1)$ correlation between the $\hat{\mathcal{W}}_i$ regions. 

\subsubsection*{\texorpdfstring{$1\rightarrow 1$}{TEXT} theorem}

For completeness, we also point out a $1\rightarrow 1$ connected wedge theorem, which follows from a simple tasks argument or from geometric observations. We consider two regions $\hat{\mathcal{C}}_1,\hat{\mathcal{R}}_1$, both in the AdS boundary, and define the scattering region,
\begin{align}
    J_{1\rightarrow 1}^\mathcal{E} = {J}^+(\mathcal{C}_1) \cap {J}^-(\mathcal{R}_1)\cap \mathcal{B},
\end{align}
and the decision region,
\begin{align}
    \hat{\mathcal{V}}_1 = \hat{J}^+(\hat{\mathcal{C}}_1)\cap \hat{J}^-(\hat{\mathcal{R}}_1).
\end{align}
By analogy with the $1\rightarrow 2$ theorem, we expect that $J_{1\rightarrow 1}^{\mathcal{E}}$ being non-empty implies the entanglement wedge of $\hat{\mathcal{V}}_1$ is brane-attached. To verify this, we can give both a tasks and geometric argument. 

From tasks, we consider an input $H^q\ket{b}_A$ at $\mathcal{C}_1$ and output $b$ at $\mathcal{R}_1$, with $q$ recorded into the brane degrees of freedom. If $J_{1\rightarrow 1}^{\mathcal{E}}$ is non-empty, then one can use a simple bulk strategy: travel to the brane, learn $q$, then send $q$ to the output point where it can be used to undo $H^q$ and recover $b$. In the bulk picture knowing $q$ is necessary to successfully recover $b$, so $\hat{\mathcal{V}}_1$ must know $q$, so $\hat{\mathcal{V}}_1$ must have the brane in its entanglement wedge.\footnote{A more rigorous argument for this would follow the strategy of section \ref{sec:QIargument}.}. 

To understand this from the geometric perspective, note that $J^+(\mathcal{C}_1)\cap J^-(\mathcal{R}_1)$ is inside the entanglement wedge of $\hat{\mathcal{V}}_1$, so $J_{1\rightarrow 1}^\mathcal{B}$ non-empty means the brane is inside the entanglement wedge of $\hat{\mathcal{V}}_1$. 

\section{Vacuum AdS\texorpdfstring{$_{2+1}$}{TEXT}}\label{sec:constantT}

In this section we give constant tension brane solutions in global AdS$_{2+1}$, then verify the connected wedge theorem by explicit calculations in that setting.
In this case the converse of Theorem \ref{thm:main} holds, and a bulk scattering region is present if and only if the entanglement wedge is connected.
This is similar to the situation for the $2\rightarrow 2$ theorem, where the converse holds for vacuum AdS$_{2+1}$ \cite{may2019quantum}.
We present a brief overview of the calculation, but relegate the details to Appendix \ref{sec:ads3-calc}.

\subsection{Constant tension branes in global AdS\texorpdfstring{$_{2+1}$}{TEXT}}

\begin{figure}
    \centering
    \subfloat[\label{fig:patch1}]{
    \tdplotsetmaincoords{15}{0}
    \begin{tikzpicture}[scale=0.9,tdplot_main_coords]
    \tdplotsetrotatedcoords{0}{30}{0}
    \draw[gray] (2,-4,0) -- (2,4,0);
    \draw[gray,dashed] (-2,-4,0) -- (-2,4,0);
    
    \begin{scope}[tdplot_rotated_coords]
    
    \draw[domain=0:180,dashed,gray] plot ({2*cos(\x+90)},{2*\x/90},{-2*sin(\x+90)});
    \draw[domain=0:180,dashed,gray] plot ({2*cos(\x+90)},{-2*\x/90},{-2*sin(\x+90)});
    
    \draw[thick,fill=black,opacity=0.5] (0,-4,2) -- (0,-4,-2) -- (0,4,-2) -- (0,4,2);
    
    \begin{scope}[canvas is xz plane at y=-4]
    \draw[domain=-90:90] plot ({2*cos(\x)}, {2*sin(\x)});
    \draw[domain=270:90,dashed,gray] plot ({2*cos(\x)}, {2*sin(\x)});
    \end{scope}
    \begin{scope}[canvas is xz plane at y=4]
    \draw[domain=-90:90] plot ({2*cos(\x)}, {2*sin(\x)});
    \draw[domain=270:90,dashed,gray] plot ({2*cos(\x)}, {2*sin(\x)});
    \end{scope}
    
    \draw[domain=0:180,thick] plot ({-2*cos(\x+90)},{2*\x/90},{-2*sin(\x+90)});
    \draw[domain=0:180,thick] plot ({-2*cos(\x+90)},{-2*\x/90},{-2*sin(\x+90)});
    
    \foreach \x in {0,...,180}
    {
        \draw[opacity=0.3,blue] ({-2*cos(\x+90)},{2*\x/90},{-2*sin(\x+90)}) -- ({-2*cos(\x+90)},{-2*\x/90},{-2*sin(\x+90)});
    }
    
    \end{scope}
    \end{tikzpicture}
    }
    \hfill
    \subfloat[\label{fig:patch2}]{
    \tdplotsetmaincoords{15}{0}
    \begin{tikzpicture}[scale=0.9,tdplot_main_coords]
    \tdplotsetrotatedcoords{0}{30}{0}
    \draw[gray] (2,-4,0) -- (2,4,0);
    \draw[gray,dashed] (-2,-4,0) -- (-2,4,0);
    
    \begin{scope}[tdplot_rotated_coords]
    
    \draw[dashed,domain=90:180,gray] plot ({2*cos(\x)},{2*\x/90},{2*sin(\x)});
    \draw[dashed,domain=-90:-180,gray] plot ({2*cos(\x)},{-2*\x/90},{2*sin(\x)});
    
    \draw[dashed,domain=90:180,gray] plot ({2*cos(\x)},{-2*\x/90},{2*sin(\x)});
    \draw[dashed,domain=-90:-180,gray] plot ({2*cos(\x)},{2*\x/90},{2*sin(\x)});
    
    \draw[thick,fill=black,opacity=0.5] (0,-4,2) -- (0,-4,-2) -- (0,4,-2) -- (0,4,2);
    
    \begin{scope}[canvas is xz plane at y=-4]
    \draw[domain=-90:90] plot ({2*cos(\x)}, {2*sin(\x)});
    \draw[domain=270:90,dashed,gray] plot ({2*cos(\x)}, {2*sin(\x)});
    \end{scope}
    \begin{scope}[canvas is xz plane at y=4]
    \draw[domain=-90:90] plot ({2*cos(\x)}, {2*sin(\x)});
    \draw[domain=270:90,dashed,gray] plot ({2*cos(\x)}, {2*sin(\x)});
    \end{scope}
    
    \draw[domain=0:90,thick] plot ({2*cos(\x)},{2*\x/90},{2*sin(\x)});
    \draw[domain=0:-90,thick] plot ({2*cos(\x)},{-2*\x/90},{2*sin(\x)});
    
    \draw[domain=0:90,thick] plot ({2*cos(\x)},{-2*\x/90},{2*sin(\x)});
    \draw[domain=0:-90,thick] plot ({2*cos(\x)},{2*\x/90},{2*sin(\x)});

    \foreach \x in {0,...,90}
    {
        \draw[opacity=0.3,blue] ({2*cos(\x)},{2*\x/90},{2*sin(\x)}) -- ({2*cos(\x)},{-2*\x/90},{2*sin(\x)});
    }
    
    \foreach \x in {0,...,-90}
    {
        \draw[opacity=0.3,blue] ({2*cos(\x)},{-2*\x/90},{2*sin(\x)}) -- ({2*cos(\x)},{2*\x/90},{2*sin(\x)});
    }

    \end{scope}
    \end{tikzpicture}
    }
    
    \caption{Global AdS$_{2+1}$ with a ETW brane. We've shown the $T=0$ case for simplicity. Poincar\'e patches are shaded in blue. (a) An edge centered choice of Poinar\'e patch. In the associated Poincar\'e solution the ETW brane is flat, described by equation \ref{eq:planarsolution}. (b) A Poincar\'e patch centered at $\sigma=0$. In Poincar\'e coordinates the brane trajectory is a hyperbola, described by equation \ref{eq:hyperbolicbrane}.}
    \label{fig:poincarepatches}
\end{figure}
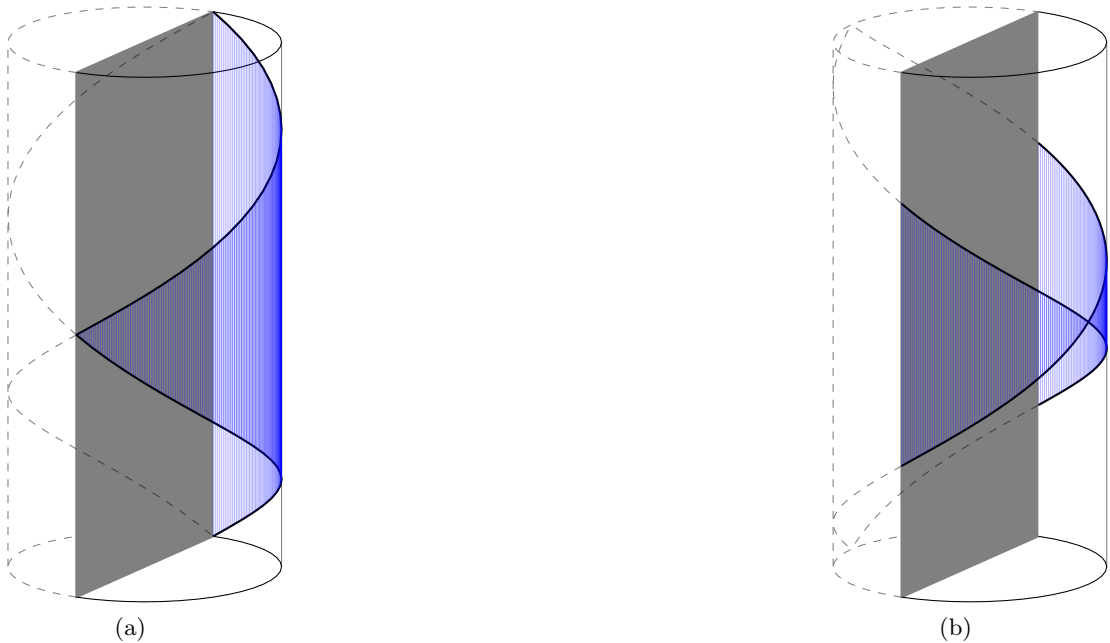

We will consider a simple model where the bulk matter action is set to zero, and the brane has constant tension. This corresponds to a Lagrangian
\begin{align}
    L_{\text{matter}}^B = -\frac{1}{8\pi G_N} T.
\end{align}
Extremizing the action \ref{eq:ETWaction}, we obtain the vacuum Einstein's equations in the bulk and a boundary condition for the brane:
\begin{align}\label{eq:constantTboundarycondition}
    (K_{ab}-Kh_{ab}) = -T h_{ab}\;.
\end{align}
We can solve this along with Einstein's equations. The solutions of interest are described by the metric
\begin{align}
    ds^2_{2+1} = \cosh^2\rho \,ds^2_{1+1} + d\rho^2 = \cosh^2\rho \left(\ell^2\frac{-d\nu^2+d\sigma}{\cos^2\sigma} \right) + d\rho^2,
    \label{eq:ds2+1}
\end{align}
with $ds_{1+1}^2$ the line element for a global $1+1$ dimensional AdS space. Allowing $-\infty<\rho<\infty$, this is global AdS$_{2+1}$. To add an ETW brane we restrict to $\rho_0<\rho<\infty$, where the brane is located at $\rho=\rho_0$ and
\begin{align}
    T = \frac{1}{\ell}\tanh (\rho_0/\ell).
\end{align}
We will call the $(\nu,\sigma, \rho)$ coordinates \emph{slicing coordinates}, since $\rho$ foliates AdS$_{1+1}$ slices to form an $_{2+1}$ spacetime.

\begin{figure}
    \centering
    \subfloat[\label{fig:angle}]{
    \tdplotsetmaincoords{25}{0}
    \begin{tikzpicture}[scale=0.7,tdplot_main_coords]
    \tdplotsetrotatedcoords{0}{25}{0}

    \begin{scope}[tdplot_rotated_coords]
    
    \draw[thick, gray,domain=65:180,<->] plot ({-1+cos(\x)},{0},{sin(\x)});
    \node at (-2,0.2,1) {$\Theta$};

    \draw[thick] (-5,0,0) -- (-1,0,0);
    
    \draw[thick] (-5,2,0) -- (-1,2,0);
    \draw[thick] (-5,-2,0) -- (-1,-2,0);
    
    \draw[thick,blue] (-1,2,0) -- (-1,-2,0);
    
    \draw[gray,thick] (-1,2,0) -- (1,2,3);
    \draw[gray,thick] (-1,0,0) -- (1,0,3);
    \draw[gray,thick] (-1,-2,0) -- (1,-2,3);
    
    \end{scope}
    
    \end{tikzpicture}
    }
    \subfloat[\label{fig:hyperbola}]{
    \tdplotsetmaincoords{15}{0}
    \begin{tikzpicture}[scale=0.7,tdplot_main_coords]
    \tdplotsetrotatedcoords{0}{25}{0}

    \begin{scope}[tdplot_rotated_coords]
    
    \draw[domain=-2.8:2.8,red,thick] plot ({\x,\x,1});
    \draw[domain=-2.8:2.8,red,thick] plot ({\x,-\x,1});
    
    \draw[domain=1:3] plot ({\x,sqrt(\x^2-1),0});
    \draw[domain=1:3] plot ({\x,-sqrt(\x^2-1),0});
    
    \draw[domain=1:3] plot ({-\x,sqrt(\x^2-1),0});
    \draw[domain=1:3] plot ({-\x,-sqrt(\x^2-1),0});
    
    \begin{scope}[canvas is xz plane at y=0]
    \draw [gray,thick,domain=0:180] plot ({1*cos(\x)}, {1*sin(\x)});
    \end{scope}
    
    \begin{scope}[canvas is xz plane at y=2.82]
    \draw [gray,thick,domain=0:180] plot ({3*cos(\x)}, {3*sin(\x)});
    \end{scope}
    
    \begin{scope}[canvas is xz plane at y=-2.82]
    \draw [gray,thick,domain=0:180] plot ({3*cos(\x)}, {3*sin(\x)});
    \end{scope}
    
    \begin{scope}[canvas is yz plane at x=0]
    \draw [gray,thick,domain=1:3] plot ({sqrt(\x^2-1)}, {\x});
    \draw [gray,thick,domain=1:3] plot ({-sqrt(\x^2-1)}, {\x});
    \end{scope}
    
    \draw [gray,thick,domain=1:3] plot ({\x/1.41}, {sqrt(\x^2-1)},{\x/1.41});
    \draw [gray,thick,domain=1:3] plot ({\x/1.41}, {-sqrt(\x^2-1)},{\x/1.41});
    
    \draw [gray,thick,domain=1:3] plot ({-\x/1.41}, {sqrt(\x^2-1)},{\x/1.41});
    \draw [gray,thick,domain=1:3] plot ({-\x/1.41}, {-sqrt(\x^2-1)},{\x/1.41});
    
    \draw[thick] (-5,0,0) -- (-1,0,0);
    \draw[blue] plot [mark=*, mark size=1.5] coordinates{(-1,0,0)};
    
    \draw[thick] (5,0,0) -- (1,0,0);
    \draw[blue] plot [mark=*, mark size=1.5] coordinates{(1,0,0)};
    
    \end{scope}
    
    \end{tikzpicture}
    }
    \caption{(a) Poincar\'e-AdS$_{2+1}$ with a constant tension ETW brane, as obtained by taking an edge-centered patch of the global spacetime, as shown in figure \ref{fig:patch2}. (b) Poincar\'e-AdS$_{2+1}$ with a constant tension ETW brane, as obtained by taking a patch as shown in figure \ref{fig:patch1}. The brane forms a hyperbola, and the two edge trajectories are $x=\pm \sqrt{1+t^2}$. The horizons $\sigma =\pm \nu$ chosen in the global geometry map to $x=\pm t$, $z=(1-\sin \Theta)/\cos\Theta$ in Poincar\'e coordinates, which we've shown in red. }
    \label{fig:poincaresols}
\end{figure}
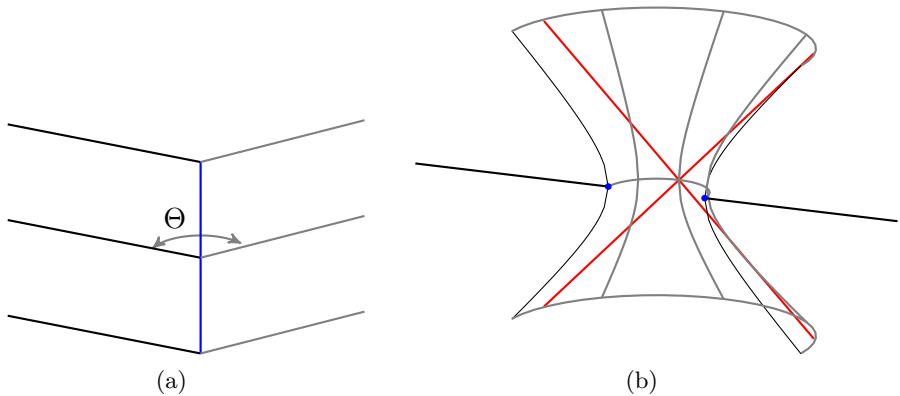

There are two ways of taking Poincar\'e patches of this spacetime that will be of interest to us. First, as shown in figure \ref{fig:patch1}, we can center our Poincar\'e patch on one of the edges, $\sigma= \pm\pi/2$. The associated Poincar\'e coordinates are related to slicing coordinates by
\begin{align}
    t = \frac{\sin \nu}{\cos \nu - \sin \sigma}, \,\,\,\,\,\, x = \frac{\cos \sigma \tanh \rho}{\cos \nu - \sin \sigma}, \,\,\,\,\,\,
    z = \frac{\cos \sigma\sech \rho}{\cos \nu - \sin \sigma}.
\end{align}
Under this transformation the boundary becomes the half line $x>0$, with one edge located at $x=0$. The other edge is at $x=\infty$. The ETW branes trajectory is  
\begin{align}\label{eq:planarsolution}
\frac{x}{z} = \tan \Theta,
\end{align}
where $\Theta$ is related to the tension $T$ by $\ell T=\sin \Theta$. Solutions of this form are shown in figure \ref{fig:angle}. 

Using this planar solution we can relate the bulk parameter $T$ to CFT data. In the CFT, one can calculate the entropy of an interval of size $L$ ending on the CFT-boundary,
\begin{align}\label{eq:boundaryentropy}
    S(L) = \frac{c_{\text{bulk}}}{6}\log \frac{L}{\epsilon} + \log g_{\mathcal{B}}.
\end{align}
The second term is known as the \emph{boundary entropy} \cite{Affleck1991}, and counts the degrees of freedom located at the edge. The Ryu-Takayanagi prescription reproduces this entropy expression in the simple constant tension model if we relate the tension and boundary entropy according to
\begin{align}
    \log g_{\mathcal{B}} = \frac{\ell}{4G_N}\text{arctanh}(\ell T).
    \label{eq:gB-tan}
\end{align}

The second Poincar\'e patch we will be interested in is centered at $\sigma=0$, as shown in figure \ref{fig:patch2}. This coordinate change is most easily performed using the embedding space formalism, see appendix \ref{sec:coords}. The ETW brane in this Poincar\'e patch is described by
\begin{align}\label{eq:hyperbolicbrane}
    x^2-t^2 +(z+\tan\Theta)^2=\sec^2\Theta
\end{align}
The edge trajectories are described by $t = \pm \sqrt{x^2-1}$. This solution was studied in \cite{Rozali:2019day} in the context of brane models of black holes and island formation, which we will also take up in section \ref{sec:islands}. A solution of this type is shown in figure \ref{fig:hyperbola}. 

\subsection{Null rays and entanglement}

In the solutions \ref{eq:ds2+1}, we will check the theorem in the case that the input and output regions are points, $\hat{\mathcal{C}}_1 = \{c_1\}$ and $\hat{\mathcal{R}}_i = \{r_i\}$.
We will calculate the travel time of null rays in the geometry (\ref{eq:ds2+1}), used to perform the bulk local strategy, and compare to a calculation of entanglement entropy on the field theory side.

We can transform slicing coordinates (\ref{eq:ds2+1}) into the following form, as discussed in appendix \ref{sec:ads3-calc}: 
\begin{equation}
    \D s_{2+1}^3 = \frac{\cosh^2 \rho}{\sin^2\theta} (-\D \nu^2 + \D \theta^2+ \sin^2\theta \, \D \varphi^2)\;, \label{eq:ads3}
\end{equation}
where $\ell_\AdS = 1$, $\theta = \sigma + \pi/2 \in [0, \pi]$, and $\varphi \in [\varphi_\mathcal{B}, \pi]$ is a warping coordinate for the copies of $\AdS_{1+1}$, with $\varphi = 0$ the position of the asymptotic region and $\varphi = \varphi_\mathcal{B}$ the location of the brane.
While the brane has the geometry of a copy of global AdS$_{1+1}$, the bulk is conformally equivalent to a patch of $\bbR\times \bbS^2$ enclosed by two lines of longitude.
In these coordinates, it is easy to trace out light cones.
As discussed in more detail in appendix \ref{sec:light-rays}, if Alice sends a signal from $c_1 = (0, \theta_0)$ light rays will arrive at the brane at angle $\theta$ at a time
\begin{equation}
    \cos [\nu(\theta)] = \cos\theta_0 \cos\theta + \sin\theta_0 \sin\theta \cos\varphi_\mathcal{B}\;. \label{eq:brane-time}
\end{equation}

Next we study the von Neumann entropy of subregions of the CFT.
This can obtained using the replica trick in the BCFT.
We start by analytically continuing the Lorentzian metric (\ref{eq:ads3}) to Euclidean time $\tau = i\nu$, and choose a defining factor to obtain the BCFT on $M_E = \bbR \times [0, \pi]$.
Following the calculation of \cite{sully2020bcft}, and as detailed in the appendix, we can calculate the entanglement entropy of the (Euclidean) interval $A := [w_1, w_2]$, for $w_j = \tau_j + i\theta_j$. The phase transition occurs at
\begin{equation}
    g_{\mathcal{B}}^{12/c} = \left|\frac{\cosh(\Delta\tau) - \cos(\Delta\theta)}{2\sin(\theta_1)\sin(\theta_2)}\right|\;,
    \label{eq:euc-cross}
\end{equation}
where $\Delta w = w_2 - w_1 = \Delta \tau + i\Delta\theta$, $\Delta \theta = \theta_2 - \theta_1$ and $\Delta\tau = \tau_2 - \tau_1$, $c$ is the central charge of the CFT, and $g_{\mathcal{B}} := \langle 0|B\rangle$ is the boundary entropy.

Returning to Lorentzian time, $\nu = -i\tau$,
(\ref{eq:euc-cross}) gives
\begin{equation}
    g_{\mathcal{B}}^{12/c} = \left|\frac{\sin[(\Delta\theta + \Delta \nu)2]\sin[(\Delta\theta - \Delta \nu)2]}{\sin(\theta_1)\sin(\theta_2)}\right|\;.\label{eq:lor-cross}
\end{equation}
We note that the brane angle $\varphi_\mathcal{B}$ is related to the boundary entropy by
\begin{equation}
    g_{\mathcal{B}}^{6/c} = \tan \left(\frac{\varphi_\mathcal{B}}{2}\right)\;.
    \label{eq:varB-g}
\end{equation}
This follows from (\ref{eq:gB-tan}) and the relation $c = 3\ell_\AdS/2G_N$, and is discussed further in appendix \ref{sec:ads3-calc}. 
In the next section, we combine these facts about light rays and entanglement to confirm the connected wedge theorem for pure $\AdS_{2+1}$ ended by constant tension branes.

\subsection{A check of the connected wedge theorem}

Let $c_1 = (0, \theta_0)$ be the input point. Without loss of generality, consider output points $r_0, r_\pi$ on opposite edges.\footnote{If they are on the same edge, $\hat{V}_1$ intersect the edge and the theorem is trivially true.} The backward light cones for these points intersect at some point $x = (\nu_1, \theta_1)$, and hence the decision region is $\hat{\mathcal{V}}_1 = \hat{J}^+(c_1) \cap \hat{J}^-(x)$.
If $\hat{\mathcal{V}}_1$ intersects the edges of the BCFT, then the boundary local strategy can be trivially performed: Alice travels to the edge, decodes her qubit, and sends the results to $r_0$ and $r_\pi$.

We will be interested in the case where this strategy cannot be performed, and hence $\hat{\mathcal{V}}_1 = \hat{D}[A]$ for a boundary interval $A$ with endpoints
\begin{align}
    L = (\theta_L, \nu_L) & = 
\frac{1}{2}(\theta_1 + \theta_0 - \nu_g, \theta_0 - \theta_1 + \nu_g), \\
R = (\theta_R, \nu_R) & = 
\frac{1}{2}(\theta_1 + \theta_0 + \nu_g, \theta_1 - \theta_0 + \nu_g)\;.
    \label{eq:endpoint-R}
\end{align}
To perform a bulk local strategy, Alice must shoot null rays in the bulk so they intersect the brane in the past of the point on the brane with boundary coordinates $x$. This strategy marginally succeeds when the light ray hits $x$ itself.\footnote{From (\ref{eq:ads3}), we note that each $\AdS_{1+1}$ slice is conformally equivalent to the flat boundary. Since this conformal factor is invisible to light rays, the point of intersection on the brane has the same boundary coordinates as the intersection on the boundary.}
If she sends it from $c_1$, (\ref{eq:brane-time}) tells us it arrives at the brane at a time $\nu_{\mathcal{B}}$ obeying
\begin{equation}
    \cos \nu_{\mathcal{B}} = \cos\theta_0 \cos\theta_1 + \sin\theta_0 \sin\theta_1 \cos\varphi_\mathcal{B}\;. \label{eq:bulk-doable}
\end{equation}
Our connected wedge theorem states that when this ray can arrive at $x$, or $\nu_{\mathcal{B}} \leq \nu_1$, $A$ has a brane-connected entanglement wedge.

Using equations (\ref{eq:lor-cross})--(\ref{eq:endpoint-R}), the transition to a brane-connected entanglement wedge occurs at a time $t_g$ obeying
\begin{equation}
     \tan^2\left(\frac{\varphi_\mathcal{B}}{2}\right) =\left|
        \frac{\cos \nu_g + \cos(\theta_1 - \theta_0)}{\cos \nu_g - \cos(\theta_1 - \theta_0)}
    \right|\;.
\end{equation}
Fixing $\theta_0, \theta_1$ and solving for $\nu_1$, some algebra shows it obeys (\ref{eq:bulk-doable}).
In other words, the transition to a connected entanglement wedge occurs precisely when the bulk local strategy becomes possible.
This explicitly verifies the connected wedge theorem for vacuum AdS$_{2+1}$.

As a simple illustration take $\theta_0 = \theta_1 = \pi/2$, corresponding to edge output points at equal times.
From (\ref{eq:varB-g}), the phase transition occurs at a time $\nu_g$ given by
\[
\tan^2\left(\frac{\varphi_\mathcal{B}}{2}\right) = \left|\frac{\sin^2(\nu_g/2)}{\sin[(\pi+\nu_g)/2]\sin[(\pi-\nu_g)/2]}\right| = \tan^2\left(\frac{\nu_g}{2}\right)\;,
\]
in other words, when $\nu_P = \varphi_\mathcal{B}$.
But from (\ref{eq:brane-time}), a light ray from $c_1$ arrives at the brane at time $\nu_{\mathcal{B}} = \varphi_\mathcal{B}$.
So the phase transition occurs precisely when Alice is able to perform the quantum task using the bulk local strategy.

\section{The connected wedge theorem and islands}\label{sec:islands}

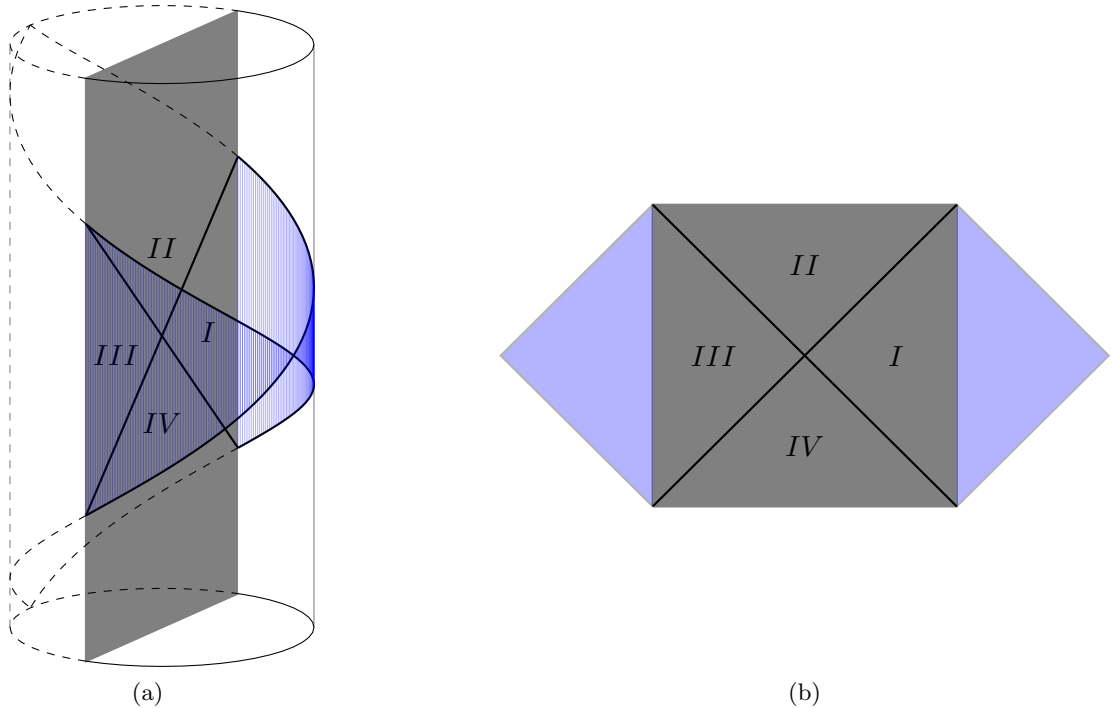
\begin{figure}
    \centering
    \subfloat[\label{fig:patch1withhorizons}]{
    \tdplotsetmaincoords{15}{0}
    \begin{tikzpicture}[scale=1,tdplot_main_coords]
    \tdplotsetrotatedcoords{0}{30}{0}
    \draw[gray] (2,-4,0) -- (2,4,0);
    \draw[gray,dashed] (-2,-4,0) -- (-2,4,0);
    
    \begin{scope}[tdplot_rotated_coords]
    
    \draw[dashed,domain=90:180] plot ({2*cos(\x)},{2*\x/90},{2*sin(\x)});
    \draw[dashed,domain=-90:-180] plot ({2*cos(\x)},{-2*\x/90},{2*sin(\x)});
    
    \draw[dashed,domain=90:180] plot ({2*cos(\x)},{-2*\x/90},{2*sin(\x)});
    \draw[dashed,domain=-90:-180] plot ({2*cos(\x)},{2*\x/90},{2*sin(\x)});
    
    \draw[thick,fill=black,opacity=0.5] (0,-4,2) -- (0,-4,-2) -- (0,4,-2) -- (0,4,2);
    
    \draw[thick] (0,-2,2) -- (0,2,-2);
    \draw[thick] (0,2,2) -- (0,-2,-2);
    
    \begin{scope}[canvas is xz plane at y=-4]
    \draw[domain=-90:90] plot ({2*cos(\x)}, {2*sin(\x)});
    \draw[domain=270:90,dashed] plot ({2*cos(\x)}, {2*sin(\x)});
    \end{scope}
    \begin{scope}[canvas is xz plane at y=4]
    \draw[domain=-90:90] plot ({2*cos(\x)}, {2*sin(\x)});
    \draw[domain=270:90,dashed] plot ({2*cos(\x)}, {2*sin(\x)});
    \end{scope}
    
    \draw[domain=0:90,thick] plot ({2*cos(\x)},{2*\x/90},{2*sin(\x)});
    \draw[domain=0:-90,thick] plot ({2*cos(\x)},{-2*\x/90},{2*sin(\x)});
    
    \draw[domain=0:90,thick] plot ({2*cos(\x)},{-2*\x/90},{2*sin(\x)});
    \draw[domain=0:-90,thick] plot ({2*cos(\x)},{2*\x/90},{2*sin(\x)});

    \node at (0.7,{3.25-3.1},0) {$I$};
    \node at (-0.7,{2.75-3.1},0) {$III$};
    \node at (0,{4.4-3.2},0) {$II$};
    \node at (0,{2-3.2},0) {$IV$};

    \foreach \x in {0,...,90}
    {
        \draw[opacity=0.3,blue] ({2*cos(\x)},{2*\x/90},{2*sin(\x)}) -- ({2*cos(\x)},{-2*\x/90},{2*sin(\x)});
    }
    
    \foreach \x in {0,...,-90}
    {
        \draw[opacity=0.3,blue] ({2*cos(\x)},{-2*\x/90},{2*sin(\x)}) -- ({2*cos(\x)},{2*\x/90},{2*sin(\x)});
    }

    \end{scope}
    \end{tikzpicture}
    }
    \hfill
    \subfloat[\label{fig:patch1boundaryview}]{
    \begin{tikzpicture}[scale=0.8]
    
    \draw[thick,fill=black,opacity=0.5] (-2.5,0) -- (2.5,0) -- (2.5,5) -- (-2.5,5) -- (-2.5,0);
    \draw[thick,opacity=0.3,fill=blue] (-2.5,0) -- (-5,2.5) -- (-2.5,5);
    \draw[thick,opacity=0.3,fill=blue] (2.5,0) -- (5,2.5) -- (2.5,5);
    
    \draw[thick] (-2.5,0) -- (2.5,5);
    \draw[thick] (2.5,0) -- (-2.5,5);
    
    \node at (0,-2.5) {$ $};
    
    \node at (1.5,2.5) {$I$};
    \node at (-1.5,2.5) {$III$};
    \node at (0,4) {$II$};
    \node at (0,1) {$IV$};
    
    \end{tikzpicture}
    }
    
    \caption{Choosing an appropriate Poincar\'{e} patch of the global spacetime, we find a two-sided black hole geometry (on the brane) coupled to two flat regions (wedges of the CFT). The end points of the two flat regions are coupled in the global picture. Note that we are most interested in the case where $T\approx 1$ and gravity localizes to the brane. We have drawn the $T=0$ case however to simplify the diagram.}
    \label{fig:penrose1}
\end{figure}

In this section, we point out that, in brane models, the $1\rightarrow 2$ connected wedge theorem reveals the formation of islands in the Ryu-Takayanagi formula. Indeed from the perspective of physics on the brane, the RT surface attaching to the brane corresponds to the formation of an island \cite{Rozali:2019day, Almheiri2020, Almheiri2020b,almheiri2020entanglement,geng2020information,geng2020massive,chen2020quantumPart1,chen2020quantumPart2}. The connected wedge theorem then relates the formation of this island to causal features of the higher dimensional AdS geometry. In this section we make more precise how we can view the brane as a black hole and a portion of the CFT as the radiation system, and finally apply the connected wedge theorem in this context. 

\subsection{The black hole and the radiation system}

We will focus on the solutions described in section \ref{sec:constantT}, which have a constant tension brane ending a pure, global, AdS$_{2+1}$ spacetime. We are most interested in the case where $T\approx 1$, where gravity localizes to the brane \cite{randall1999alternative}. As noted in the introduction, choosing $\hat{\mathcal{C}}_1,\hat{\mathcal{R}}_1,\hat{\mathcal{R}}_2$ to be extended regions gives no additional power to the connected wedge theorem in these solutions, and consequently for simplicity we will take $\hat{\mathcal{C}}_1=c_1$, $\hat{\mathcal{R}}_1=r_1$, $\hat{\mathcal{R}_2}=r_2$ where $c_1,r_1,r_2$ are points on the boundary of AdS, and in particular $r_1, r_2$ sit in the edge. 

For constant tension solutions we have two simplifications that will prove useful in understanding the connected wedge theorems relationship to islands. The first simplification is that for constant tension branes light rays run tangent to the brane. This allows us to define horizons in the brane by choosing points $r_1$ and $r_2$ on the edge, and considering their forward light cones, 
\begin{align}
    H_1 &= [\partial J^+(r_1)]_\mathcal{B}, \nonumber \\
    H_2 &= [\partial J^+(r_2)]_\mathcal{B}.
\end{align}
These horizons intersect at $p_\mathcal{B}=(\sigma=0,\nu=0)$. From these horizons, define regions $I-IV$ as in figure \ref{fig:penrose1}. Region $II$ is the black hole interior, while $I$ and $IV$ are the right and left exteriors. 

The second simplification is that the $1\rightarrow 2$ theorem is if and only if for constant tension solutions. This will let us conclude that an island forms if and only if a certain scattering configuration occurs. This is not essential, as we may still be interested in a sufficient condition for the formation of an island. 

To make the black hole features of these constant tension brane solutions more explicit, consider going to the Poincar\'e patch shown in figure \ref{fig:patch1withhorizons}. This patch includes the entire black hole, along with two wedge shaped portions of the CFT and a portion of the AdS bulk. Forgetting the bulk picture and focusing on the brane coupled to CFT picture, we have the spacetime shown in figure \ref{fig:patch1boundaryview}. 

Explicitly the Poincar\'e patch is described by a metric
\begin{align}
    ds^2 = \frac{\ell^2}{z^2}(-dt^2+dx^2+dz^2)
\end{align}
with brane located at
\begin{align}
    x^2-t^2 +(z+\tan\Theta)^2=\sec^2\Theta,
\end{align}
where $\Theta$ is related to the tension $T$ according to $T=\sin \Theta$. The Poincar\'e patch includes only the $-\pi/2<\nu<\pi/2$ portion of the brane. The points $r_1$ and $r_2$ are mapped to $x=t=-\infty$ and $-x=t=-\infty$. The details of this coordinate change are given in appendix \ref{sec:coords}.

In Poincar\'e coordinates the edge trajectory is $x = \pm \sqrt{1+t^2}$. These trajectories asymptote to the light rays $x=\pm t$. Mapping the horizons $v=\pm \sigma$ to Poincar\'e we find horizons
\begin{align}
    z = \frac{1-\sin\Theta}{\cos\Theta} \,\,\,\,\,\,\,,\,\,\,\,\,\, x= \pm t .
\end{align}
One can also verify directly in the Poincar\'e geometry that these are the horizons by studying null geodesics in the brane geometry \cite{Rozali:2019day}.

Next we should identify the radiation system. The entire CFT is coupled to the black hole at the two edges, and information can escape from the black hole into anywhere in the CFT. It seems sensible however to not consider the portion of the CFT which reconstructs the black hole exterior regions as being part of the radiation system. It is straightforward to identify the CFT dual to the left and right exterior black hole regions. The interval $Y_1=\{\sigma\in (-\pi/2,0),\nu=0\}$ has region $I$ inside its entanglement wedge. Similarly the interval $Y_2=\{\sigma\in(0,\pi/2),\nu=0\}$ has region $II$ inside its entanglement wedge. This excludes $D(Y_1)$ and $D(Y_2)$ from the radiation system.  

The remaining portion of the CFT is the future and past of the point 
\begin{align}
    x = (\nu=0,\sigma=0,\rho=\infty).
\end{align}
The future of $x$ reconstructs region $II$ of the brane, so we should identify this with the radiation system. To specify that radiation has been collected only up until a certain time, we can choose a second point $c_1$ and define
\begin{align}
    \hat{R} = J^+(x) \cap J^-(c_1).
\end{align}
For $c_1$ at an early time so that $R$ is small, the entanglement wedge of $R$ will be disconnected from the brane, and $R$ does not see inside of the black hole. At late enough times though, $\mathcal{E}_W(R)$ connects to the brane. Where this transition occurs will be controlled by the connected wedge theorem. Note also that since the minimal surfaces are at constant $\sigma$, they will in fact lie exactly on the horizons. This is illustrated in figure \ref{fig:Islandappearance}. 

\subsection{The connected wedge theorem and behind the horizon}

Finally, we can apply the connected wedge theorem to this black hole on the brane. In fact, we need a time reversed variant of the theorem, which follows immediately from Theorem \ref{thm:main} (we also specialize to the case where the input and output regions are points),
\begin{theorem}\label{thm:timereversedmain}\textbf{($2\rightarrow 1$ connected wedge theorem)}
Consider three points $r_1,r_2,c_1$ in an asymptotically AdS$_{2+1}$ spacetime with an end-of-the-world brane, with $c_1$ in the boundary and $r_1,r_2$ on the edge. Then if
\begin{align}
    J_{12\rightarrow 1} = J^+(r_1) \cap J^+(r_2) \cap J^-(c_1)
\end{align}
is non-empty, the entanglement wedge of
\begin{align}
    \hat{\mathcal{V}_1} = \hat{J}^+(r_1) \cap \hat{J}^+(r_2) \cap \hat{J}^-(c_1)
\end{align}
is attached to the brane. 
\end{theorem}
The two input points of the theorem we identify with the points $r_1$ and $r_2$ we used above to define the black hole horizons $H_1$ and $H_2$. The region $\hat{\mathcal{V}}_1$ becomes the subsystem of the radiation which has been collected since $J^+(r_1)\cap J^+(r_2)=J^+(x)$, so $\hat{\mathcal{V}}_1=\hat{R}$. 

Applying Theorem \ref{thm:timereversedmain} along with its converse (which holds because we are in the constant tension solutions) gives a simple condition for when the radiation system reconstructs a portion of the black hole interior: an island forms if and only if there is a causal curve from the black hole interior into the radiation system. 

This causal picture for island formation immediately reveals a set of simple operators that probe behind the black hole horizon. In particular consider an operator $\mathcal{O}_y$, which is localized near a point $y$, with $y$ in $\hat{R}$ and in the future of the black hole interior (such points exist by our theorem). These operators directly probe the black hole interior by virtue of being in its future. 

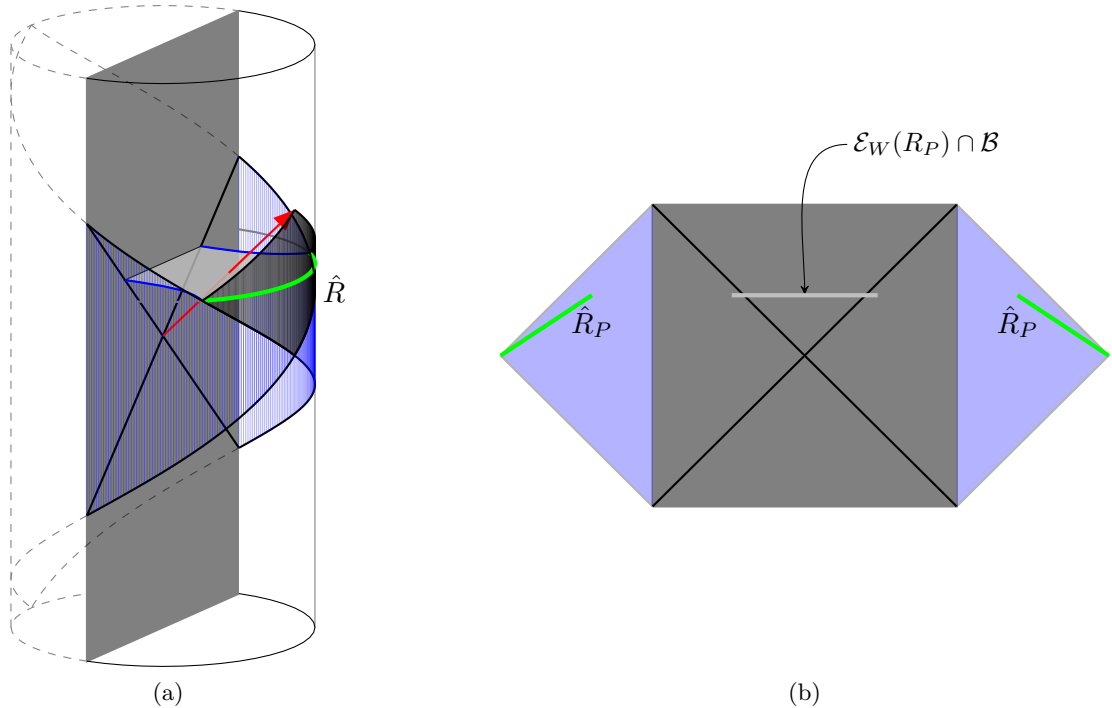
\begin{figure}
    \centering
    \subfloat[\label{fig:patch1withV}]{
    \tdplotsetmaincoords{15}{0}
    \begin{tikzpicture}[scale=1,tdplot_main_coords]
    \tdplotsetrotatedcoords{0}{30}{0}
    \draw[gray] (2,-4,0) -- (2,4,0);
    \draw[gray,dashed] (-2,-4,0) -- (-2,4,0);
    
    \begin{scope}[tdplot_rotated_coords]
    
    \draw[dashed,domain=90:180,gray] plot ({2*cos(\x)},{2*\x/90},{2*sin(\x)});
    \draw[dashed,domain=-90:-180,gray] plot ({2*cos(\x)},{-2*\x/90},{2*sin(\x)});
    
    \draw[dashed,domain=90:180,gray] plot ({2*cos(\x)},{-2*\x/90},{2*sin(\x)});
    \draw[dashed,domain=-90:-180,gray] plot ({2*cos(\x)},{2*\x/90},{2*sin(\x)});
    
    \draw[thick,fill=black,opacity=0.5] (0,-4,2) -- (0,-4,-2) -- (0,4,-2) -- (0,4,2);
    
    \draw[thick] (0,-2,2) -- (0,2,-2);
    \draw[thick] (0,2,2) -- (0,-2,-2);
    
    \draw[thick, red] (0,0,0) -- (1,1,0);
    
    \begin{scope}[canvas is xz plane at y=-4]
    \draw[domain=-90:90] plot ({2*cos(\x)}, {2*sin(\x)});
    \draw[domain=270:90,dashed,gray] plot ({2*cos(\x)}, {2*sin(\x)});
    \end{scope}
    \begin{scope}[canvas is xz plane at y=4]
    \draw[domain=-90:90] plot ({2*cos(\x)}, {2*sin(\x)});
    \draw[domain=270:90,dashed,gray] plot ({2*cos(\x)}, {2*sin(\x)});
    \end{scope}
    
    \draw[domain=0:90,thick] plot ({2*cos(\x)},{2*\x/90},{2*sin(\x)});
    \draw[domain=0:90,thick] plot ({2*cos(\x)},{-2*\x/90},{2*sin(\x)});

    \foreach \x in {0,...,90}
    {
        \draw[opacity=0.3,blue] ({2*cos(\x)},{2*\x/90},{2*sin(\x)}) -- ({2*cos(\x)},{-2*\x/90},{2*sin(\x)});
    }
    
    \node at (2.6,1,0) {$\hat{R}$};
    
    \begin{scope}[canvas is xz plane at y=1]
    \draw[domain=-90:90,thick,gray] plot ({2*cos(\x)}, {2*sin(\x)});
    \draw[domain=45:135,fill=lightgray,opacity=0.8] plot ({2*cos(\x-90)}, {2*sin(\x-90)}) -- (1.41,1.41) to [out=-135,in=0] (0,1) --  (0,-1) to [out=0,in=135] (1.41,-1.41) ;
    \draw[blue,thick] (1.41,1.41) to [out=-135,in=0] (0,1);
    \draw[blue,thick] (1.41,-1.41) to [out=135,in=0] (0,-1);
    \end{scope}
    
    \draw[thick, red,-triangle 45] (1,1,0) -- (2,2,0);
    
    \draw[domain=0:-90,thick] plot ({2*cos(\x)},{2*\x/90},{2*sin(\x)});
    \draw[domain=0:-90,thick] plot ({2*cos(\x)},{-2*\x/90},{2*sin(\x)});
    
    \foreach \x in {0,...,-90}
    {
        \draw[opacity=0.3,blue] ({2*cos(\x)},{-2*\x/90},{2*sin(\x)}) -- ({2*cos(\x)},{2*\x/90},{2*sin(\x)});
    }
    
    \foreach \x in {0,...,-90}
    {
    \draw[fill=black!50!,opacity=0.5] ({2*cos(\x/2)},{2+2*\x/180},{2*sin(\x/2)}) -- ({2*cos(\x/2)},{0-2*\x/180},{2*sin(\x/2)});
    }
    
    \foreach \x in {1,...,90}
    {
    \draw[fill=black!50!,opacity=0.5] ({2*cos(\x/2)},{2-2*\x/180},{2*sin(\x/2)}) -- ({2*cos(\x/2)},{2*\x/180},{2*sin(\x/2)});
    }
    
    \begin{scope}[canvas is xz plane at y=1]
    \draw [green,ultra thick,domain=45:135] plot ({2*cos(\x-90)}, {2*sin(\x-90)});
    \end{scope}

    \draw[thick,domain=-45:0] plot ({2*cos(\x)},{2+2*\x/90},{2*sin(\x)});
    \draw[thick,domain=45:0] plot ({2*cos(\x)},{2-2*\x/90},{2*sin(\x)});
    
    \draw[thick,domain=-45:0] plot ({2*cos(\x)},{0-2*\x/90},{2*sin(\x)});
    \draw[thick,domain=45:0] plot ({2*cos(\x)},{0+2*\x/90},{2*sin(\x)});

    \end{scope}

    \end{tikzpicture}
    }
    \hfill
    \subfloat[\label{fig:patch1boundarywithV}]{
    \begin{tikzpicture}[scale=0.8]
    
    \draw[thick,fill=black,opacity=0.5] (-2.5,0) -- (2.5,0) -- (2.5,5) -- (-2.5,5) -- (-2.5,0);
    
    \draw[thick,opacity=0.3,fill=blue] (-2.5,0) -- (-5,2.5) -- (-2.5,5);
    \draw[ultra thick,green] (-5,2.5) -- (-3.5,3.5);
    
    \draw[thick,opacity=0.3,fill=blue] (2.5,0) -- (5,2.5) -- (2.5,5);
    \draw[ultra thick,green] (5,2.5) -- (3.5,3.5);

    \node[below] at (3.5,3.5) {$\hat{R}_P$};
    \node[below] at (-3.5,3.5) {$\hat{R}_P$};
    
    \node at (2,6) {\small{$\mathcal{E}_W(R_P) \cap \mathcal{B}$}};
    
    \draw[->] (0.7,6) to [out=180,in=90] (0,3.5);
    
    \draw[thick] (-2.5,0) -- (2.5,5);
    \draw[thick] (2.5,0) -- (-2.5,5);
    
    \node at (0,-2.5) {$ $};
    
    \draw[ultra thick,lightgray] (-1.2,3.5) -- (1.2,3.5);
    
    \end{tikzpicture}
    }
    
    \caption{ (a) The radiation system $R$ (time-slice in green) picked out by the connected wedge theorem sits outside the Poincar\'e patch. (b) A nearby region $\hat{R}_P$ inside the patch has $\hat{R}_1$ inside of its domain of dependence, so that $\hat{R}_P$ has an island whenever $\hat{R}_1$ does. The entanglement wedge of $\hat{R}_P$ (shown in light gray) will include a small portion of the black hole exterior in its entanglement wedge.}
    \label{fig:penrose2}
\end{figure}

Notice that the radiation system $\hat{R}$ sits outside of the Poincar\'e patch we identified above. Thus it sits outside of the black hole spacetime. Ideally, we would understand which subregions of the Poincar\'e patch reconstruct the black hole interior. To do this, we need only note that a nearby subregion $\hat{R}_P$ of the Poincar\'e patch includes $\hat{R}$ in its domain of dependence. See figure \ref{fig:penrose2}. Evolving the state on this subregion forward using the global Hamiltonian, we can construct the state of the radiation system $\hat{R}_1$. Notice that $\hat{R}_P$ is slightly larger than $\hat{R}$ and will include a small portion of the black hole exterior in its entanglement wedge. 

To write operators which probe behind the black hole horizon in the Hilbert space of $V_P$, we can start with the operators $\mathcal{O}_y$ which live in $\hat{\mathcal{V}}_1$ and time evolve backward using the global Hamiltonian. We continue this time evolution until $\mathcal{O}_y$ is some non-local operator $\mathcal{O}_{y,P}$ living on $V_P$. 

It is interesting that time evolution with the global Hamiltonian, along with local operators, can be used to probe the black hole interior. We should perhaps be unsurprised however, as the situation is analogous to the traversable wormhole \cite{gao2017traversable}: in both cases we have a left and right CFT (or in our setting, BCFT), which we couple and then time evolve to find that information from behind the black hole horizon has emerged at the boundary. In the traversable wormhole the coupling is a double trace term which can be understood perturbatively, while in our setting the coupling is due to time evolution with the global Hamiltonian.\footnote{We thank Henry Lin for pointing out this analogy to us.} 

\section{Discussion}\label{sec:discussion}

In this paper we have proven the $1\rightarrow 2$ connected wedge theorem. The theorem is motivated by a quantum tasks argument, and proven using the focusing theorem. The tasks argument gives an operational reason why the theorem should be true: if the bulk scattering region is non-empty, the boundary CFT requires the decision region $\hat{\mathcal{V}}_1$ to know information stored on the brane. Otherwise, the CFT is unable to reproduce bulk physics. The focusing theorem based proof relies on the null membrane, a structure that allows comparison of the areas of brane-detached extremal surfaces and brane-attached surfaces. When the scattering region is non-empty, we showed there exists a null membrane that connects a brane-detached extremal surface to a brane-attached one with less area. 

Below we make a number of comments. 
\subsection{Better bounds on mutual information}

The key technical tool used here to complete the quantum tasks argument for the $1\rightarrow 2$ connected wedge theorem was Lemma \ref{lemma:mutualinfobound}, which gave a bound
\begin{align}
    \frac{1}{2}I(\hat{\mathcal{V}}_1:\bar{Q})\geq n(-\log 2^{h(2\epsilon)}\beta) - 1 +O((\epsilon/\beta)^n). \nonumber
\end{align}
This bound was first shown in \cite{may2021holographic}. It is interesting to ask if this bound can be improved further. Supposing a fraction $1-\delta$ of the $\mathbf{M}$ tasks need to be completed successfully for the $\mathbf{M}^{\times n}$ task to be declared successful, it is straightforward to achieve 
\begin{align}
    \frac{1}{2}I(\hat{\mathcal{V}}_1:\bar{Q}) = (1-\delta )n.
\end{align}
Thus, $(1-\delta)n$ is the best lower bound on the mutual information we can hope for. 
Given such a bound, and assuming we can take $\delta=O(1/n)$, we could directly find that the boundary region $\hat{\mathcal{V}}_1$ approximately reconstructs $Q$ \cite{schumacher2002entanglement,schumacher2002approximate}. With the existing bound, we can instead only conclude $I(\hat{\mathcal{V}}_1:\bar{Q})=O(1/G_N)$, then use the Ryu-Takayanagi formula to conclude this means the entangling surface are brane-anchored, then use the understanding of entanglement wedge reconstruction to conclude this means $\hat{\mathcal{V}}_1$ reconstructs $Q$. Post-hoc, we can interpret this as being due to $\hat{\mathcal{V}}_1$ needing $Q$ to undo $H^q$ and complete the task. The better bound presented above would more directly connect the task argument to bulk reconstruction. 

\subsection{Relation to correlation functions}

An un-explored question is the relationship between the quantum tasks considered here and features of CFT correlation functions. Recall from \cite{gary2009local,heemskerk2009holography,penedones2011writing,maldacena2017looking} that when there is a bulk point $p$ with $c_1,c_2\prec p \prec r_1,r_2$ and $p$ is null separated from each of the four points, there is a perturbative singularity in four point functions $\langle \mathcal{O}(c_1)\mathcal{O}(c_2)\mathcal{O}(r_1)\mathcal{O}(r_2)\rangle$. The appearance of this point $p$ also signals the appearance of a scattering region, and so the $2\rightarrow 2$ connected wedge theorem implies large mutual information between the decision regions.

The connected wedge theorem and the appearance of perturbative singularities are related by the bulk geometry --- indeed the singularity in the four point function, via the bulk point, implies a large mutual information. In \cite{may2019holographic}, the authors suggested that this should have a CFT explanation, but so far no explanation has been offered. 

This question also has a natural analogue in the $1\rightarrow 2$ connected wedge theorem. In particular it is plausible that the three point function $\langle \mathcal{O}(c_1) \psi(r_1)\psi(r_2)\rangle$, where $\psi$ is an edge operator, has a perturbative singularity when the three operator insertion points are null separated from a single point on the brane.\footnote{This is the case for the two point function $\langle \calO(c_1)\calO(r_1)\rangle$ in a BCFT. For further discussion, and a complementary perspective on causality and spectral properties arising from two-point functions in a BCFT, see the upcoming work \cite{upcoming}.} Comparing to the $1\rightarrow 2$ theorem, these singularities would then imply the entanglement wedge of the decision region is connected to the brane.

\subsection{\texorpdfstring{$1\rightarrow 2$}{TEXT} theorem in planar brane solutions}

\begin{figure}
    \centering
    \begin{tikzpicture}[scale=0.65]
    
    \draw[thick, black, fill=black!60!,opacity=0.8] (-6,0) -- (-4,2) -- (-2,0) -- (-4,-2) -- (-6,0);
    
    \draw[lightgray] (-10,8) -- (0,8) -- (0,-2) -- (-10,-2) -- (-10,8);

    \draw[fill=black] (-4,-2) circle (0.15);
    \node[below] at (-4,-2.2) {$\hat{\mathcal{C}}_1=x_1$};
    \draw[fill=black] (-4,2) circle (0.15);
    \node[above] at (-4,2.2) {$y_1$};

    \draw[thick,fill=blue,opacity=0.3] (-4,2) -- (0,6) -- (0,-2) -- (-4,2);
    \draw[thick,fill=blue,opacity=0.3] (-4,2) -- (-10,8) -- (-10,-2) -- (-8,-2) -- (-4,2);
    
    \node at (-4,0) {$\hat{\mathcal{V}}_1$};

    \node at (-1,2) {$\hat{\mathcal{R}}_1$};
    \node at (-8,2) {$\hat{\mathcal{R}}_2$};
    
    \draw[black,ultra thick] (0,-2) -- (0,8);
    
    \end{tikzpicture}
    \caption{View of the boundary of Poincar\'e-AdS$_{2+1}$. The edge is located at $x=0$. A region $\hat{\mathcal{V}}_1$ is specified, and we are interested in using the connected wedge theorem to determine if the entanglement wedge of $\hat{\mathcal{V}}_1$ is attached to the brane. The figure shows a choice of regions $\hat{\mathcal{R}}_1, \hat{\mathcal{R}}_2$ and $\hat{\mathcal{C}}_1$ which can be used in the theorem. Notice that the input region $\hat{\mathcal{C}}_1$ is taken to be a point $x_1$.}
    \label{fig:poincarebranesetup}
\end{figure}
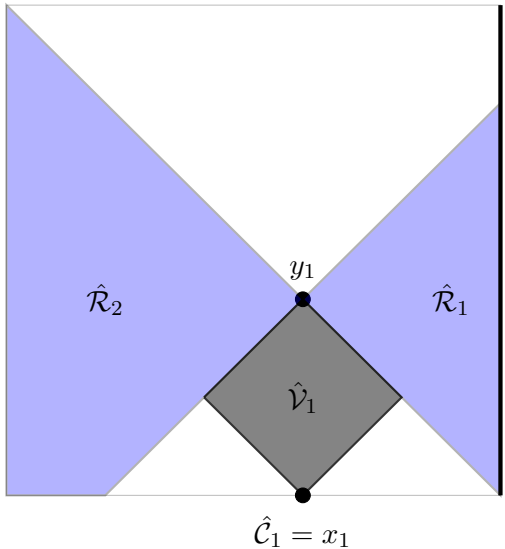

By using extended input and output regions, \cite{may2021holographic} applied the $2\rightarrow 2$ connected wedge theorem non-trivially in Poincar\'e-AdS$_{2+1}$. Here, we have mostly focused on a point based formulation, and on a class of global solutions with constant tension branes ending pure AdS. However we can also use extended input and output regions and apply the $1\rightarrow 2$ theorem non-trivially in Poincar\'e-AdS$_{2+1}$ with a brane. 

Consider in particular the pure AdS solutions with planar branes discussed in section \ref{sec:constantT}. The boundary is the half plane defined by $x<0$, and the brane sits at
\begin{align}
    x/z = \sin \Theta.
\end{align}
Suppose we are given a region on the boundary $\hat{\mathcal{V}}_1$, and we would like to apply the $1\rightarrow 2$ theorem to determine if its entanglement wedge is brane attached or detached. Call $x_1$ the earliest point on $\hat{\mathcal{V}}_1$, and $y_1$ the latest point on $\hat{\mathcal{V}}_1$ so that $\hat{\mathcal{V}}_1=\hat{J}^+(x_1)\cap \hat{J}^-(y_1)$. To apply the theorem non-trivially, choose
\begin{align}
    \hat{\mathcal{C}}_1 &= x_1 \nonumber \\
    \hat{\mathcal{R}}_1 &= D((y_1,0]) \nonumber \\
    \hat{\mathcal{R}}_2 &= D((-\infty,y_1)) 
\end{align}
where by $D(\cdot)$ we mean the domain of dependence. This is shown in figure \ref{fig:poincarebranesetup}. Then the $1\rightarrow 2$ connected wedge theorem applies non-trivially, and in fact in these solutions the entanglement scattering region will be non-empty exactly when the entanglement wedge of $\hat{\mathcal{V}}_1$ is connected. 

\subsection{Higher dimensions}

The $1\rightarrow 2$ connected wedge theorem is true for any asymptotically AdS spacetime where the bulk and brane matter satisfy the NEC. However, it is possible that in some spacetimes there will only be trivial configurations of the theorem, meaning that when the bulk scattering region is non-empty the decision region $\hat{\mathcal{V}}_1$ touches the brane, and so immediately has a connected entanglement wedge. 

We have focused on the example of asymptotically global AdS$_{2+1}$, and discussed Poincar\'e-AdS$_{2+1}$ in the last section, where there are many non-trivial configurations. It would be interesting to better understand however when the theorem applies non-trivially in higher dimensions. 

\subsection{Evaporating black holes}

In section \ref{sec:islands} we applied the connected wedge theorem to the static, two sided black hole model introduced in \cite{Rozali:2019day}. While information does escape from this black hole, it is in thermal equilibrium with the radiation system and does not evaporate. The connected wedge theorem applies much more generally however, including to models of evaporating black holes, since the theorem is proven in the context of dynamical spacetimes. It would be interesting to do this explicitly, for example in the dynamical models of \cite{Rozali:2019day}. 

\subsection*{Acknowledgements}

We thank Jon Sorce and Geoff Penington for helpful discussions. Mark Van Raamsdonk made important comments on the connection to islands discussed in section \ref{sec:islands}. Jamie Sully was involved in early discussions leading to conjecturing Theorem \ref{thm:main}. Jason Pollack provided feedback on this manuscript. AM is supported by a C-GSM award given by the National Science and Engineering Research Council of Canada.
DW is supported by an International Doctoral Fellowship from the University of British Columbia.

\appendix

\section{Details for AdS\texorpdfstring{$_{2+1}$}{TEXT} calculation} \label{sec:ads3-calc}

\subsection*{Light rays}\label{sec:light-rays}

Let's consider the reflection of bulk light rays in the simplest case,
vacuum AdS$_{2+1}$ with a brane of tension $T$.
We can write the global metric (for $\ell_\AdS = 1$)
in the slicing coordinates AdS$_{1+1}$:
\begin{align}
\D s_{2+1}^2 = \cosh^2 \rho \, \D s_{1+1}^2 + \D \rho^2 = \cosh^2 \rho \, \left(\frac{- \D \nu^2
    + \D\theta^2}{\sin^2\theta}\right) + \D \rho^2,
\end{align}
where $\rho_0$ is the position of the brane and $\rho = \infty$ the boundary, global Lorentzian time is $\nu \in \bbR$, and
$\theta = \sigma + \pi/2 \in [0, \pi]$.
Null rays are simple in conformally flat
coordinates, which we find by defining a new warping coordinate
\begin{equation}
  \label{eq:2}
  \varphi = \frac{\pi}{2} + 2 \tan^{-1} \left[\tanh
    \left(\frac{\rho}{2}\right)\right] = 2 \tan^{-1} e^\rho, \quad
  \D\varphi^2 = \frac{\D \rho^2}{\cosh^2\rho},
\end{equation}
with $\varphi \in [\varphi_\mathcal{B}, \pi]$ for a brane at $\varphi_\mathcal{B} =
\varphi(\rho_0)$.
Then our global metric becomes
\begin{equation}
  \label{eq:3}
  \D s_3^2 = \frac{\cosh^2 \rho}{\sin^2\theta}(- \D \nu^2 + \D\theta^2 +
  \sin^2\theta \, \D \varphi^2),
\end{equation}
which is conformally equivalent to patch of $\bbS^2 \times
\bbR$ enclosed by two lines of longitude.
Since null rays do not see the conformal factor, our problem reduces to
propagating light rays on the sphere.
With respect to some affine parameter $\lambda$, we have null geodesic
equation
\[
-\dot{\nu}^2 + \dot{\theta}^2 + \sin^2\theta \, \dot{\varphi}^2 = 0.
\]
If we set $\dot{\nu}=1$, our problem reduces to finding geodesic lengths
on the sphere, with affine time measuring these lengths.

A null ray will start at some initial point $\theta_0$ and with
some initial direction $\theta_0'$ at the boundary $\varphi = \pi$.
It travels into the bulk, reflects off the brane at
$\varphi_\mathcal{B}$, and finally returns to the boundary at some final
position $\theta_1$.
From the cosine rule for spherical trigonometry, the geodesic distance to the brane obeys
\begin{equation}
  \label{eq:6}
  d = \cos^{-1}[\cos\theta_0 \cos \theta_1  +\sin\theta_0 \sin\theta_1 \cos(\varphi_\mathcal{B})].
\end{equation}
Thus, the global time it takes a null ray to reach the brane with respect to the parameterisation $\dot{\nu} = 1$ is
\begin{equation}
  \label{eq:7}
  \nu(\theta) = \cos^{-1}\big[\cos\theta_0 \cos \theta  +\sin\theta_0 \sin\theta \cos(\varphi_\mathcal{B})\big]
\end{equation}

\subsection*{Entanglement entropy}\label{sec:entanglement-entropy}

We now calculate entanglement entropy from the field theory side.

First, we analytically continue $\tau = i\nu$, so that
\begin{equation}
  \label{eq:8}
  \D s_{3}^2 = \frac{\cosh^2 \rho}{\sin^2\theta}(\D\tau^2 + \D\theta^2 + \sin^2\theta \, \D\varphi^2).
\end{equation}
Choosing a defining function to remove the prefactor as we approach
the boundary, the dual BCFT is defined on $[0, \pi] \times \bbR$.
The first step is to map the strip to the upper half-plane (UHP), $z = x
+ iy$ for $y \geq 0$.
Let $w = \tau + i\theta = \log z$.
Then correlation functions for primary operators $\calO_i$ on the strip and UHP are related by
\begin{equation}
  \label{eq:10}
  \langle\calO_1(w_1)\cdots \calO_k(w_k)\rangle_\text{strip} = \prod_i |z_i|^{\Delta_i}\langle\calO_1(z_1)\cdots \calO_k(z_k)\rangle_\text{UHP}.
\end{equation}
We define the distances $z_{ij} = |z_i - z_j|$ and $z_{i\bar{j}} = |z_j - \bar{z}_j|$ for future convenience.

A twist operator creates an $n$-fold branched cover of the geometry
via boundary conditions.
The one-point function for a twist in the presence of a boundary is
\begin{equation}
  \label{eq:15}
  \langle\Phi_n(w_1)\rangle_\text{strip} = \left|\frac{z_1}{z_{1\bar{1}}}\right|^{d_n}g_{\mathcal{B}}^{1-n},
\end{equation}
where $g_{\mathcal{B}} := \langle 0|B\rangle$ is the boundary entropy \cite{Affleck1991}, and the
twist scaling dimension for central charge $c$ and replica number $n$ is given by \cite{Calabrese2016}
\begin{equation}
  \label{eq:13}\quad d_n = \frac{c}{12}\left(n-\frac{1}{n}\right).
\end{equation}
A gap and small OPE coefficients \cite{sully2020bcft} imply the simple form for
a correlator of twists:
\begin{equation}
  \label{eq:11}
  \langle \Phi_n(w_1)\Phi_{-n}(w_2)\rangle_\text{strip} =
  \min\left\{\left|\frac{z_1z_2}{z_{12}^2}\right|^{d_n}, \left|\frac{z_1z_2}{z_{1\bar{1}}z_{2\bar{2}}}\right|^{d_n}g_{\mathcal{B}}^{2(1-n)}\right\}.
\end{equation}
The entanglement entropy is given by the limit
\begin{align}
  \label{eq:14}
  S_{w_1 w_2} & = \lim_{n\to 1^+}\frac{1}{1-n} \log \langle
                \Phi_n(w_1)\Phi_{-n}(w_2)\rangle_\text{strip} \notag
  \\
  & = \min\left\{ \frac{c}{6}\log
    \left|\frac{z_{12}^2}{z_1z_2}\right|, \frac{c}{6}\log
    \left|\frac{z_{1\bar{1}}z_{2\bar{2}}}{z_1z_2}\right| + 2\log g_{\mathcal{B}}\right\}.
\end{align}
We have neglected the UV regulator, since it cancels when we calculate the transition between expressions.
This occurs at
\begin{equation}
    g_{\mathcal{B}}^{12/c} =
    \left|\frac{z_{12}^2}{z_{1\bar{1}}z_{1\bar{2}}}\right| = \left|\frac{\cosh(\Delta\tau) - \cos(\Delta\theta)}{2\sin(\theta_1)\sin(\theta_2)}\right|
\end{equation}
where $w_j = \tau_j + i\theta_j$, $\Delta \tau = \tau_2 - \tau_1$, and $\Delta \theta = \theta_2 - \theta_1$.
Reverting to $\nu = -i\tau$, this becomes
\begin{equation}
    g_{\mathcal{B}}^{12/c} = \left|\frac{\sin[(\Delta\theta+\Delta \nu)/2]\sin[(\Delta\theta-\Delta \nu)/2]}{\sin(\theta_1)\sin(\theta_2)}\right|.
    \label{eq:guess}
\end{equation}

\subsection*{Connected wedge in AdS\texorpdfstring{$_{2+1}$}{TEXT}}

To relate the location of the brane in different coordinates, first note that
\[
\rho_0 = \frac{6}{c}\log g_{\mathcal{B}}\;.
\]
Hence, by (\ref{eq:2}),
\begin{equation}
    g_{\mathcal{B}}^{12/c} = \tan^2\left(\frac{\varphi_{\mathcal{B}}}{2}\right)\;.
    \label{eq:varphi-tg}
\end{equation}
Consider an input point $c_1 = (\theta_0, 0)$, and two edge output points $r_1 = (0, t_1)$, $r_2 = (\pi, t_2)$.
The backward light cones intersect at coordinates
\begin{equation}
    x = \frac{1}{2}(\nu_1 - \nu_2+\pi, \nu_1 + \nu_2 - \pi) = (\theta_1, \nu_g).
\end{equation}
Similarly, the forward light cone of $c_1$ and the backward cone of $x$ intersect at two points,
\begin{align}
L = (\theta_L, \nu_L) & = 
\frac{1}{2}(\theta_1 + \theta_0 - \nu_g, \theta_0 - \theta_1 + \nu_g) \\
R = (\theta_R, \nu_R) & = 
\frac{1}{2}(\theta_1 + \theta_0 + \nu_g, \theta_1 - \theta_0 + \nu_g)\;.
\end{align}

In order to successfully use a bulk strategy, Alice must send a bulk light ray so that it hits the brane in the past of the point on the brane with boundary coordinates $x$.
The extreme case is when her null ray hits $x$ itself.
From (\ref{eq:6}), this occurs at a boundary time $\nu_{\mathcal{B}}$ given by
\begin{equation}
    \cos(\nu_{\mathcal{B}}) = \cos\theta_0 \cos\theta_1 + \sin\theta_0 \sin\theta_1 \cos(\varphi_\mathcal{B}).
    \label{eq:bounce}
\end{equation}
We expect that this is precisely the time at which $(L,R)$ experiences a phase transition in entanglement entropy.
From (\ref{eq:guess}), the transition occurs at
\begin{align}
    g^{12/c} = \tan^2\left(\frac{\varphi_\mathcal{B}}{2}\right) & =
    \left|\frac{\sin[(\nu_g+\theta_1-\theta_0)/2]\sin[(\nu_g+\theta_0-\theta_1)/2]}{\sin[(\theta_0+\theta_1-\nu_g)/2]\sin[(\theta_0+\theta_1-\nu_g)/2]}\right|
    \notag \\
    & = \left|
        \frac{\cos \nu_g + \cos(\theta_1 - \theta_0)}{\cos \nu_g - \cos(\theta_1 - \theta_0)}
    \right|\;,
    \label{eq:transition2}
\end{align}
where we have simplified with trigonometric identities.
To verify the connected wedge theorem, we will show from (\ref{eq:bounce}) and (\ref{eq:transition2}) that $\nu_g = \nu_{\mathcal{B}}$.
We first use the trigonometric identity
\begin{equation}
    \tan^2\left(\frac{\varphi_\mathcal{B}}{2}\right) = \frac{1 - \cos(\varphi_\mathcal{B})}{1 + \cos(\varphi_\mathcal{B})}\;.
    \label{eq:tan2}
\end{equation}
We can isolate $\cos(\varphi_\mathcal{B})$ in (\ref{eq:bounce}). Substituting this expression into (\ref{eq:tan2}) yields
\begin{align}
 \tan^2\left(\frac{\varphi_\mathcal{B}}{2}\right)
    & = \frac{\sin\theta_0\sin\theta_1 + \cos\theta_0\cos\theta_1 - \cos \nu_{\mathcal{B}}}{\sin\theta_0\sin\theta_1 - \cos\theta_0\cos\theta_1 + \cos \nu_{\mathcal{B}}} = \left|
        \frac{\cos \nu_{\mathcal{B}} + \cos(\theta_1 - \theta_0)}{\cos \nu_{\mathcal{B}} - \cos(\theta_1 - \theta_0)}
    \right|\;.
\end{align}
Comparing to (\ref{eq:transition2}), we find $\nu_{\mathcal{B}} = \nu_g$ as claimed.

\section{Coordinate systems and embedding space} \label{sec:coords}

Here, we briefly discuss the different coordinate systems used for AdS.
Rather than explicitly map between coordinates, we use the embedding space formalism, following \cite{Karch_2020} closely.
Recall that we can view $\AdS_{d+1}$ as (the universal cover of) the hyperboloid in $\mathbb{R}^{2,d-1}$, given by
\begin{equation}
    X_0^2 + X_{d+1}^2 - \sum_{i=1}^d X_i^2 = \ell_\AdS^2\;.\notag
\end{equation}
We set $\ell_\AdS = 1$ for convenience.
Different choices of coordinates map to different parametrizations of this hyperboloid.
For instance, consider standard global coordinates on $\AdS_{d+1}$:
\begin{equation}
    \D s^2_{d+1} = -\cosh^2 \hat{\rho} \, \D \hat{t} + \D \hat{\rho}^2 + \sinh^2 \hat{\rho} \, \D \Omega_{d-1}^2\;, \notag
\end{equation}
where $\hat{t}$ is global (Lorentzian) time, and the $\Omega_i$ are spherical coordinates on $\mathbb{S}^{d-1}$.
This corresponds to the parametrization
\begin{align}
    X_0 & = \cosh \hat{\rho} \cos \hat{t}\notag \\
    X_i & = \Omega_i \sinh \hat{\rho}  \notag\\
    X_{d+1} & = \cosh \hat{\rho} \sin \hat{t}\;.\notag
\end{align}
In this paper, we employ the slicing coordinates
\begin{align}
    \D s_{d+1}^2 &= \cosh^2 \rho\, \D s_d^2 + \D \rho^2\;,\notag
\end{align}
with global coordinates $\D s_d^2$ on the slices. This arises from the parametrization
\begin{align}
    X_0 & = \cosh \rho \, \cosh r \cos \nu \notag\\
    X_a & = \Omega_a \cosh \rho \sinh r \notag \\
    X_d & = \sinh \rho \notag\\
    X_{d+1} & = \cosh \rho \, \cosh r \sin \nu \;,\notag
\end{align}
where $a = 1, \ldots, d-1$ correspond to spherical coordinates for $\mathbb{S}^{d-2}$ on the $\AdS_d$ slices.
In the global coordinates on $\AdS_{d+1}$ or $\AdS_d$, we can always compactify to the "Einstein static universe" coordinates $\hat{\sigma}, \sigma$ defined by
$$
\tan \hat{\sigma} = \sinh \hat{\rho}, \quad \tan \sigma = \sinh \rho,
$$
with (for instance)
\begin{align}
\D s_{d}^2 &= \frac{-d\nu^2 + d\sigma^2 + \sin^2\theta \, d\Omega_{d-2}^2}{\cos^2\sigma}.
\end{align}
Note that for $d > 2$, $\rho > 0$, and hence $ \sigma \in [0, \pi/2)$.
However, for $d= 2$, we can trade in the 0-sphere $\Omega^0 = \{\pm 1\}$ and take $\rho \in\mathbb{R}$, hence $\sigma \in [-\pi/2, \pi/2)$.
This is the coordinate system used in (\ref{eq:ds2+1}).
Finally, there is the Poincar\'{e} slicing,
\begin{equation}
    \D s^2_{d+1} = \frac{1}{z^2}(-\D t^2 + \D x^2 + x^2 \D \Omega_{d-2}^2)\;,\notag
\end{equation}
with parametrization
\begin{align}
    X_0 & = \frac{1}{2z}(z^2 + x^2 - t^2 + 1) \notag\\
    X_a & = \frac{x\Omega_a}{z} \notag\\
    X_d & = \frac{1}{2z}(z^2 + x^2 - t^2 - 1)\notag \\
    X_{d+1} & = \frac{t}{z}\;.\notag
\end{align}
For a brane at fixed $\rho = \rho_0$, setting $X_d$ equal in these three parametrizations leads to
\begin{equation}
    \sinh \rho_0 = \Omega_d \,\sinh \hat{\rho} = \frac{1}{2z}(z^2 + x^2 - t^2 - 1)\;.
\end{equation}

\bibliographystyle{jhep}
\bibliography{biblio}

\end{document}